\documentclass[english,twocolumn,prd,superscriptaddress,nofootinbib,floatfix,showpacs]{revtex4-1}

\usepackage{float,graphicx}
\usepackage[T1]{fontenc}
\usepackage{amsmath}
\usepackage{amssymb}
\usepackage{pstricks}
\usepackage{graphicx}
\usepackage{latexsym}
\usepackage{epsfig}
\usepackage{amssymb}
\usepackage{amsmath}
\usepackage{pstricks}
\usepackage{graphicx}
\usepackage{hyperref}

\begin{document}

\title{The Vainshtein mechanism beyond the quasi-static approximation}

\author{Hans~A.~Winther}
\email[Email address: ]{hans.winther@astro.ox.ac.uk}
\affiliation{Astrophysics, University of Oxford, DWB, Keble Road, Oxford, OX1 3RH, UK}

\author{Pedro~G.~Ferreira}
\email[Email address: ]{p.ferreira1@physics.ox.ac.uk  }
\affiliation{Astrophysics, University of Oxford, DWB, Keble Road, Oxford, OX1 3RH, UK}

\date{\today}

\begin{abstract}
Theories of modified gravity, in both the linear and fully non-linear regime, are often studied under the assumption that the evolution of the new (often scalar) degree of freedom present in the theory is quasi-static. This approximation significantly simplifies the study of the theory, and one often has good reason to believe that it should hold. Nevertheless it is a crucial assumption that should be explicitly checked whenever possible. In this paper we do so for the Vainshtein mechanism. By solving for the full spatial and time evolution of the Dvali-Gabadadze-Porrati and the Cubic Galileon model, in a spherical symmetric spacetime, we are able to demonstrate that the Vainshtein solution is a stable attractor and forms no matter what initial conditions we take for the scalar field. Furthermore,the quasi-static approximation is also found to be a very good approximation whenever it exists. For the best-fit Cubic Galileon model, however, we find that for deep voids at late times, the numerical solution blows up at the same time as the quasi-static solution ceases to exist. We argue that this phenomenon is a true instability of the model.
\end{abstract}


\maketitle

\section{Introduction}

There is overwhelming evidence that the Universe is undergoing accelerated expansion. A possible, popular, explanation is the existence of a cosmological constant $\Lambda$. The  $\Lambda$ Cold Dark Matter ($\Lambda$CDM) model is, thus far in perfect agreement with observations. An alternative is that new dynamical degrees of freedom must be at play in our Universe, in the form of a new fluid that has been dubbed dark energy \cite{2006IJMPD..15.1753C}. Yet another alternative is that gravity is modified on cosmological scales \cite{2012PhR...513....1C}.

General Relativity has been exquisitely tested in the solar system \cite{2015arXiv150107274B}. Any theory that attempts to modify gravity must conform to the stringent limits coming from these experiments; this usually requires the presence of a screening mechanism \cite{2004PhRvD..69d4026K,2008arXiv0805.3430M,2010PhRvL.104w1301H,2013arXiv1312.2006K}, i.e.  a way of dynamically hiding modifications of gravity in regions where we have tested general relativity to great precision. The Vainshtein mechanism \cite{vainshtein_paper} is one such mechanism and relies on non-linear differential operators to screen the fifth-force in high density regions of spacetime. This mechanism is responsible for, the viability of popular modified gravity models such as massive gravity \cite{2014LRR....17....7D}, DGP \cite{2000PhLB..485..208D} and the Galileon \cite{2009PhRvD..79h4003D,2009PhRvD..79f4036N,2011PhRvD..84f4039D}. 
 
The study of modified gravity models often require us to solve complicated, non-linear, partial differential equations for the new degrees of freedom. For scales that are small compared to the horizon one can often apply what is known as the {\it quasi-static} approximation to significantly simplify the analysis. This approximation roughly means that we neglect most terms involving time-derivatives of perturbed quantities relative to those containing spatial derivatives.
On linear, sub-horizon, scales it has been shown that, for a wide class of models, we can safely assume the quasi-static approximation. However as was shown in \cite{2014PhRvD..89b3521N} its is possible to have large deviations on scales close to the horizon. In general we have that the quasi-static approximation breaks down for scales larger than the sound horizon of the dynamical degree(s) of freedom in question \cite{2015arXiv150306831S}.

In the non-linear regime of structure formation the main tool to obtain accurate predictions is {\it N}-body simulations. Over the last decade there have been a number of studies of screening mechanisms \cite{2013JCAP...10..027B,2013JCAP...11..012L,2012JCAP...10..002B,2013JCAP...04..029B,2014A&A...562A..78L,2014arXiv1403.6492W,2012ApJ...748...61D,2014JCAP...07..058F,2009PhRvD..80d3001S,2009PhRvD..80l3003S} in the non-linear regime by performing such simulations. Most of these simulations have assumed the quasi-static approximation. 
Recently a handful of papers have appeared where scalar fields have been allowed to evolve dynamically \cite{2013PhRvL.110p1101L,2014PhRvD..89h4023L,2015JCAP...02..034B,2014PhRvD..90l3521B,2015arXiv150407142H} and treated beyond the quasi-static approximation. These studies have demonstrated that, in most situations of cosmological interest, the approximation is very good. Nevertheless interesting non-static effect have also been found in certain models, such as, for example, the formation of domain walls \cite{2013PhRvL.110p1101L}. 

For the Vainshtein mechanism, which is the topic of this paper, the quasi-static approximation has been shown to work well for linear perturbations \cite{2012PhRvD..86l4016B}, but for the non-linear regime the only study we are aware of looking at non-static effects was in \cite{2009PhRvD..80d3001S} for the Dvali-Gabadadze-Porratti (DGP) model. It was shown that the quasi-static approximation is a self-consistent approximation in the sense that by assuming it and solving the approximate equations, one could then calculate the effect of the neglected terms and show that they were indeed negligible. 

Another interesting class of models with Vainshtein screening are the so-called Galileon models (see e.g. \cite{2010PhRvD..82b4011G,2009PhRvD..80b4037C,2010PhRvD..82j3015A,2010PhRvL.105k1301D,2010PhRvD..82l4054N,2012JCAP...03..043A,2013JCAP...11..056B,2013PhRvD..87j3511B,2009PhRvD..80b4037C,2012JCAP...03..043A,2010PhRvD..82l4054N,2012PhRvD..86l4016B,2014JCAP...04..029B}). The Cubic Galileon model, being the simplest Galileon, has been found to be a viable alternative to $\Lambda$CDM for explaining cosmic acceleration\footnote{There are some issues related to the ISW effect in the Galileon models that might be is in tension with observations.} \cite{2014arXiv1406.0485B,2014PhRvD..90b3528B}. However, the best-fit Cubic Galileon model has a peculiar property: if we study the theory in the quasi-static approximation for a spherical symmetric body in a Friedman Robertson Walker metric then we find that, for some configurations (depending on the model parameters and the object in question), the analytic field-profile ceases to exist. The same effect has been found in {\it N}-body simulations of this model. It has been speculated \cite{2013JCAP...10..027B} that this might be an artifact of the quasi-static approximation and that if non-static effects are taken into account then this problem could be solved. This is one of the questions we aim to answer here.

In this paper we present spherical symmetric cosmological simulations of the DGP and the Cubic Galileon model beyond the quasi-static approximation. We want to answer questions like: does the Vainshtein mechanism always form? Is the quasi-static approximation always valid in the non-linear regime? Are other approximations widely used in simulations of such models, like neglecting terms with metric potentials whenever they do not have a $\nabla^2$ in front of them, also valid? In all the questions above we naively expect the answer to be yes. Our aim is to put these questions to rest and place the quasi-static approximation on a firm footing by demonstrating and quantifying its validity in the non-linear regime.

The setup of this paper is as follows. In Sec.~(\ref{sect:review}) we introduce the DGP and the Cubic Galileon model, in Sec.~(\ref{sect:vainshtein}) we discuss the Vainshtein mechanism and the quasi-static approximation. In Sec.~(\ref{sect:beyondqs}) we present the equations needed to go beyond the quasi-static approximation and in Sec.~(\ref{sect:numberical}) we discuss the numerical implementation before presenting code-tests in Sec.~(\ref{sec:codetest}) and finally the results of our simulations in presented in Sec.~(\ref{sect:results}).

Unless stated otherwise we always work in units of $c = \hbar = 1$ and $M_{\rm Pl} \equiv \sqrt{\hbar c/ (8\pi G)} = 1$. Unless stated otherwise we use the metric sign-convention $(-,+,+,+)$.

\section{The Vainshtein Mechanism}\label{sect:review}

The Vainshtein Mechanism is the name of the mechanism with which the Galileon and DGP model is able to evade the stringent local gravity constraints in the solar-system. For a comprehensive review of how the mechanism (and other screening mechanisms) works in detail see \cite{2013arXiv1312.2006K}. We will  focus on how it arises in the cosmological context and in particular its relation with the quasi-static limit. In this section we review two simple models that have the Vainshtein mechanism.

\subsection{The DGP model}

The DGP model \cite{2000PhLB..485..208D} is a modified gravity model where we are confined to live in a four-dimensional brane, embedded in a five-dimensional spacetime. The gravitational action is given by
\begin{align}\label{eq:dgp}
S_{\rm DGP} = \int {\rm d}^4x \sqrt{-g} \frac{R^{(4)}}{16\pi G} + \int {\rm d}^5x \sqrt{-g} \frac{R^{(5)}}{16\pi G^{(5)}},
\end{align}
where $^{(4)}$ and $^{(5)}$ denote quantities on the brane and in the bulk respectively. The relative sizes of the two gravitational constants is a parameter of the model known as the crossover scale, $r_c$,
\begin{align}\label{eq:rc}
r_c = \frac{1}{2}\frac{G^{(5)}}{G},
\end{align}
For scales smaller than $r_c$ gravity looks four-dimensional. This model, in its simplest form, has two branches of solutions. The first branch, the {\it normal branch}, requires dark energy to produce cosmic acceleration of the Universe. The second branch, the {\it self-accelerating branch}, is able to produce acceleration without a cosmological constant  but is in tension with observations \cite{2006PhLB..642..432F} and is plagued by theoretical problems associated with the propagation of ghosts \cite{2003JHEP...09..029L}. For our purposes, i.e. to test the validity of the quasi-static approximation, these problems are not important and we will look at both branches.

Assuming a homogeneous and isotropic matter distribution, an empty Minkowski bulk and a (spatially) flat brane leads to the Friedmann equation
\begin{align}
H^2 \pm \frac{H}{r_c} = \frac{8\pi G}{3}\rho_m
\end{align}
where the sign $\pm$ refers to the two branches mentioned above. In the accelerating branch we obtain
\begin{align}
\frac{H(a)}{H_0} = \sqrt{\Omega_{rc}} + \sqrt{\Omega_{rc} + \Omega_m a^{-3}}
\end{align}
where $\Omega_{rc} = \frac{1}{4r_c^2H_0^2}$ and $\Omega_m = 1 - \frac{1}{r_cH_0}$. Looking at perturbations about the Friedmann solution, an additional degree of freedom appears which is associated with displacement of the brane. This is the so-called brane-bending mode $\phi$ and is determined by the field equation \cite{2007PhRvD..75h4040K,2003JHEP...09..029L,2004JHEP...06..059N}
\begin{align}
\square\phi  + \frac{r_c^2}{3\beta}\left[(\square\phi)^2 - (\nabla_{\mu}\nabla_{\nu}\phi)^2\right] = \frac{8\pi G}{3\beta}\delta\rho
\end{align}
where
\begin{align}
\beta(a) = 1 \pm 2H(a)\left(1 + \frac{\dot{H}}{3H^2(a)}\right)
\end{align}
Since $\phi$ is a perturbative quantity the background value of this is simply $\overline{\phi} = 0$. The gravitational potential $\Psi$ in the model is the sum of the standard Newtonian potential $\Psi_N$ and the brane-bending mode,
\begin{align}
\Psi = \Psi_N + \frac{\phi}{2}
\end{align}
giving rise to a fifth-force on matter-particles $F_\phi = \frac{1}{2}\nabla\phi$.

In the case of the normal branch ($+$ sign in $\beta$) we will in this paper simply assume that the background evolution of the model is the same as in $\Lambda$CDM. 

\subsection{The Cublic Galileon}
The Cublic Galileon model also containis a Vainshtein mechanism and the resulting equations are very similar to that of DGP. However there are some important differences as we shall see below. The action is given by\footnote{Note that in this subsection, and only in this subsection, we use the metric sign-convention $(+,-,-,-)$ to be consistent with the convention used in \cite{2013JCAP...10..027B}.} 
\begin{align}
S = \int d^4x\sqrt{-g}\left[\frac{R}{2} - \frac{c_2}{2}\mathcal{L}_2 - \frac{c_3}{2}\mathcal{L}_3 - \mathcal{L}_{\rm matter}\right]
\end{align}
where $c_2,c_3$ are dimensionless parameters,
\begin{align}
\mathcal{L}_2 &= (\nabla\phi)^2\\
\mathcal{L}_3 &= \frac{2}{M^3}(\nabla\phi)^2\square\phi
\end{align}
and  $M^3 = M_{\rm Pl}H_0^2$.  The Einstein equations following from the action are
\begin{align}
G_{\mu\nu} &= T^{\rm matter}_{\mu\nu} +  T^{\phi(2)}_{\mu\nu} + T^{\phi(3)}_{\mu\nu}
\end{align}
where the energy-momentum tensor of the scalar field is given by
\begin{align}
T^{(2)\phi}_{\mu\nu} =& c_2\left[\nabla_{\mu}\phi\nabla_{\nu}\phi - \frac{1}{2}g_{\mu\nu}(\nabla\phi)^2\right]\\
T^{(3)\phi}_{\mu\nu} =& \frac{2c_3}{M^3}\left[\square\phi\nabla_{\mu}\phi\nabla_{\nu}\phi + g_{\mu\nu}\nabla_{\alpha}\phi\nabla_{\beta}\phi \nabla^{\alpha}\nabla^{\beta}\phi\right.\nonumber\\
&- 2\nabla_{\lambda} \phi\nabla_{(\mu}\phi\nabla_{\nu)} \nabla^{\lambda}\phi \big]
\end{align}
A variation of the action with respect to $\phi$ gives us a modified Klein-Gordon equation:
\begin{align}\label{kg_full}
\square\phi + \frac{2c_3}{c_2M^3}\left[(\square\phi)^2 - (\nabla_{\mu}\nabla_{\nu}\phi)^2 - R_{\mu\nu}\nabla^{\mu}\phi\nabla^{\nu}\phi\right] = 0
\end{align}
In a flat Friedmann-Robertson-Lemaitre-Walker metric (FRLW) the Einstein equations gives us the Friedmann equations
\begin{align}\label{hubbeleqs}
3H^2 &= \overline{\rho}_m + \frac{1}{2}c_2 \dot{\overline{\phi}}^2 + 6c_3\frac{H \dot{\overline{\phi}}^3}{H_0^2}\\
H^2+\dot{H} &= -\frac{1}{6}\left[ \overline{\rho}_m + 2c_2\dot{\overline{\phi}}^2 + 6c_3\frac{\dot{\overline{\phi}}^2}{H_0^3}(H\dot{\overline{\phi}} - \ddot{\overline{\phi}})\right]
\end{align}
and the equation of motion reduces to
\begin{align}\label{eombackground}
\ddot{\overline{\phi}}+3H\dot{\overline{\phi}} + \frac{6c_3}{c_2H_0^2}(3H^2\dot{\overline{\phi}}^2 + 2\dot{\overline{\phi}}\ddot{\overline{\phi}}H + \dot{\overline{\phi}}^2\dot{H}) = 0
\end{align}
These equations have a late time de Sitter attractor given by $H\dot{\overline{\phi}} = -\frac{c_2}{6c_3}H_0^2$. To obtain the correct dark energy density today we need
\begin{align}
\Omega_{\phi 0} = \frac{c_2\dot{\overline{\phi}_0}^2}{6H_0^2} + 2c_3\frac{\dot{\overline{\phi}_0}^3}{H_0^3}
\end{align}
and by evaluating this expression using the attractor we find $\Omega_{\phi 0} = 1 - \Omega_m = -\frac{c_2^3}{6^3c_3^2}$. The value of $c_2$ can be fixed by perfoming a rescaling of $\phi$ so we can without loss of generality choose\footnote{The sign here is not determined by the rescaling. We choose '$-$' which is the same as in the best-fit cosmological model found in \cite{2013JCAP...10..027B}.} $c_2 = -1$. We then choose
\begin{align}
c_3 = \frac{1}{6\sqrt{6(1-\Omega_{m})}}
\end{align}
 so the model has the same number of free parameters as $\Lambda$CDM. In the following we will use the best-fit parameters from the analysis in \cite{2013JCAP...10..027B} which can be translated into $c_2=-1$ and $c_3 = 0.08$ corresponding to $\Omega_m = 0.277$ and $\Omega_{\phi 0} = 0.723$.

In Sec.~(\ref{sect:beyondqs}) we will discuss the specific form of the equations above when we go to a spherical symmetric spacetime.

\subsection{The Vainshtein Mechanism in the Quasi-Static Approximation}\label{sect:vainshtein}

The quasi-static approximation can be defined as the limit of the theory for which time derivatives of the perturbations around a given background can be neglected. Roughly this reduces to making the replacement $\dot{\phi}\to \dot{\overline{\phi}}$ where an overline denotes a background quantity. For cosmological perturbation theory we also neglect terms proportional to $\dot{\Phi}$ where $\Phi$ is any metric potential. The discussion below will be for the Cubic Galileon, but the same equations with some small modifications, also apply for the DGP model.

Applying the quasi-static approximation to the Klein Gordon equation Eq.~(\ref{kg_full}) for the Cubic Galileon we find 
\begin{align}\label{full_kg_qs}
\nabla^2\phi+ \frac{1}{3\beta_1H_0^2a^2}\left[(\nabla^2\phi)^2 - (\nabla_i\nabla_j\phi)^2\right] =  \frac{a^2}{3\beta_2}\delta\rho_m
\end{align}
where
\begin{align}
\beta_1(a) &= \frac{1}{6c_3}\left[-c_2 - 4c_3(\ddot{\overline{\phi}}+2H\dot{\overline{\phi}}) + 2c_3^2\dot{\overline{\phi}}^4\right]\\
\beta_2(a) &= \frac{2H_0^2}{\dot{\overline{\phi}}^2}\beta_1
\end{align}
are functions of the scale factor. The DGP model fits also into this description by taking $\beta_1(a) = \beta(a)/(r_cH_0)^2$ and $\beta_2(a) = \beta(a)$.

The above equations are general, but if we assume spherical symmetry then we can simplify further as it is integrable; integrating it over $r$ we obtain
\begin{align}
\frac{\phi'}{r} + \frac{2}{3\beta_1H_0^2a^2} \left(\frac{\phi'}{r}\right)^2 =  \frac{a^2}{12\pi\beta_2}\frac{\delta M(r)}{r^3}
\end{align}
where $\delta M(r) = 4\pi\int_0^r\delta\rho_m r^2 dr$. This is an algebraic equation in $\phi'/r$ which can be solved to give us
\begin{align}
\frac{\phi'}{r} &= \frac{3\beta_1H_0^2a^2}{4}\left[ \sqrt{1 + \frac{16}{9 \beta_1\beta_2 H_0^2}\frac{G\delta M(r)}{r^3}}  - 1\right]\nonumber\\
&= \frac{3\beta_1H_0^2a^2}{4}\left[ \sqrt{1 + \frac{r_V^3}{r^3}}-1\right]
\end{align}
where $r_V^3 = \frac{8 r_S}{9\beta_1\beta_2 H_0^2}$ is the  {\it Vainshtein radius} of an object with Schwarchild radius\footnote{The mass $\delta M$ is measured relative to the cosmic mean so a void-like structure has a negative $r_S$.} $r_S = 2G \delta M$. This equation can be rewritten as
\begin{align}\label{vsteinsol}
\phi' = \alpha \times \frac{2r^3}{r_V^3}\left[ \sqrt{1 + \frac{r_V^3}{r^3}}-1\right]\times \frac{G\delta M(r)}{r^2}
\end{align}
with $\alpha = \frac{2a^2}{3\beta_2}$. 

If we now take, see Eq.~(\ref{poisson_eq_full}), $\nabla^2\Psi \simeq \nabla^2\Psi_N - \frac{c_3\dot{\phi}^2}{H_0^2}\nabla^2\phi$, we have that the gravitational force $F = \nabla\Psi$ in the Newtonian limit predicted by the Galileon model can be rewritten as $F = F_N + F_{\phi} = \frac{G_{\rm eff}}{G} F_N$ where $F_N$ is the Newtonian expression for the force and where the effective gravitational constant, which is a function of both time and scale, satisfies
\begin{align}
\frac{G_{\rm eff}}{G}  = 1 - \frac{c_3\dot{\overline{\phi}}^2}{3\beta_1} \times \frac{2r^3}{r_V^3}\left[ \sqrt{1 + \frac{r_V^3}{r^3}}-1\right]
\end{align}
The equivalent expression for the DGP model can be recovered by making the replacement $-\frac{c_3\dot{\overline{\phi}}^2}{3\beta_1} \to \frac{1}{3\beta(a)}$. Outside the Vainshtein radius $r \gg r_V$ gravity is modified as
\begin{align}
\frac{G_{\rm eff}-G}{G}  \simeq \sigma
\end{align}
where $\sigma = \frac{c_3}{3(-\beta_1)}$ for the Cubic Galileon and $\sigma = \frac{1}{3\beta(a)}$ for the DGP model. Within the Vainshtein radius on the other hand ($r \ll r_V$) we have
\begin{align}
\frac{G_{\rm eff}-G}{G}  \simeq \sigma \times \left(\frac{r}{r_V}\right)^{3/2} \ll \sigma
\end{align}
If an object it massive enough as to have a large Vainshtein radius then,  for $r \ll r_V$ General Relativity is recovered.

\subsection{Breakdown of the quasi-static Vainshtein solution in the Cubic Galileon model}\label{sect:breakdown}

\begin{figure}
\includegraphics[width=1.0\columnwidth]{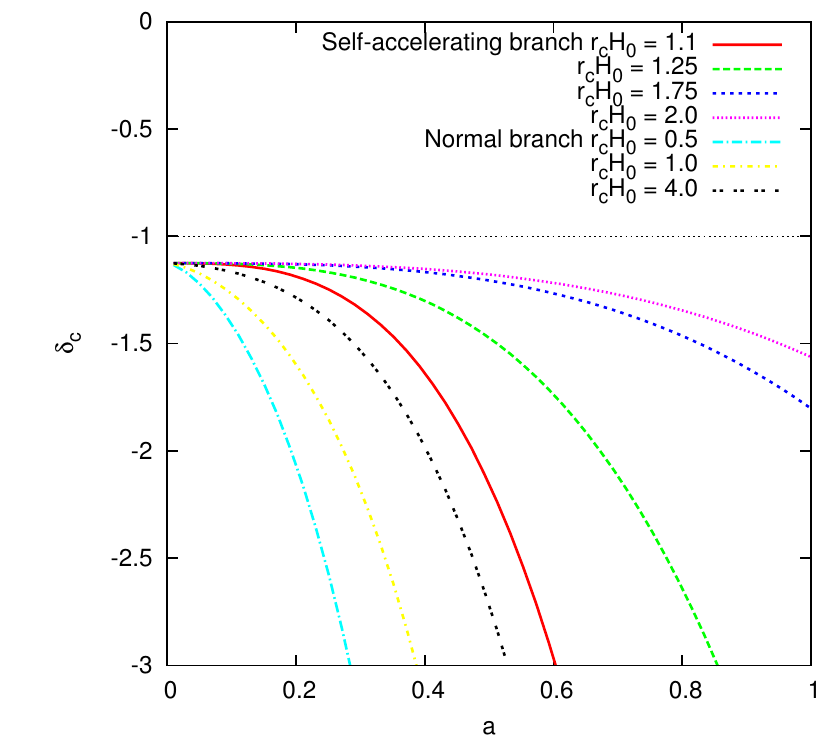}
\caption{The allowed density contrast $\delta_c$ of a void versus scale factor $a$ for which the condition Eq.~(\ref{breakdown_cond}) is satisfied for the DGP model.}
\label{fig:denconditionDGP}
\end{figure}

The quasi-static spherical symmetric solution derived in the previous subsection is only valid when
\begin{align}
1+ \frac{r_V^3}{r^3} \geq 0
\end{align}
otherwise we get a complex solution for $\phi'$. This condition translates into
\begin{align}
\frac{9\beta_1\beta_2 a^3}{8\Omega_m} \geq -\frac{3\int_0^r\delta_m r^2 dr}{r^3} \equiv -\left<\delta_m\right>
\end{align}
For overdensities this condition is always satisfied\footnote{This is only for the Cubic and Quartic Galileon. For the general Galileon model this problem can also occur for $\mathcal{O}(1)$ overdensities as shown in \cite{2013JCAP...11..056B}.} as $\delta_m > 0$ and $\beta_1\beta_2 \propto \left(\beta_1/\dot{\overline{\phi}}\right)^2 \geq 0$. For DGP we also have $\beta_1\beta_2 = \beta^2(a)/(r_cH_0)^2 \geq 0$. 
Furthermore, in both the self-accelerating and the normal branch DGP model we have that the critical value for the density contrast $\delta_c \leq - \frac{9}{8} < -1$ at all times, see Fig.~(\ref{fig:denconditionDGP}), so the quasi-static solution always exists.

\begin{figure}
\includegraphics[width=1.0\columnwidth]{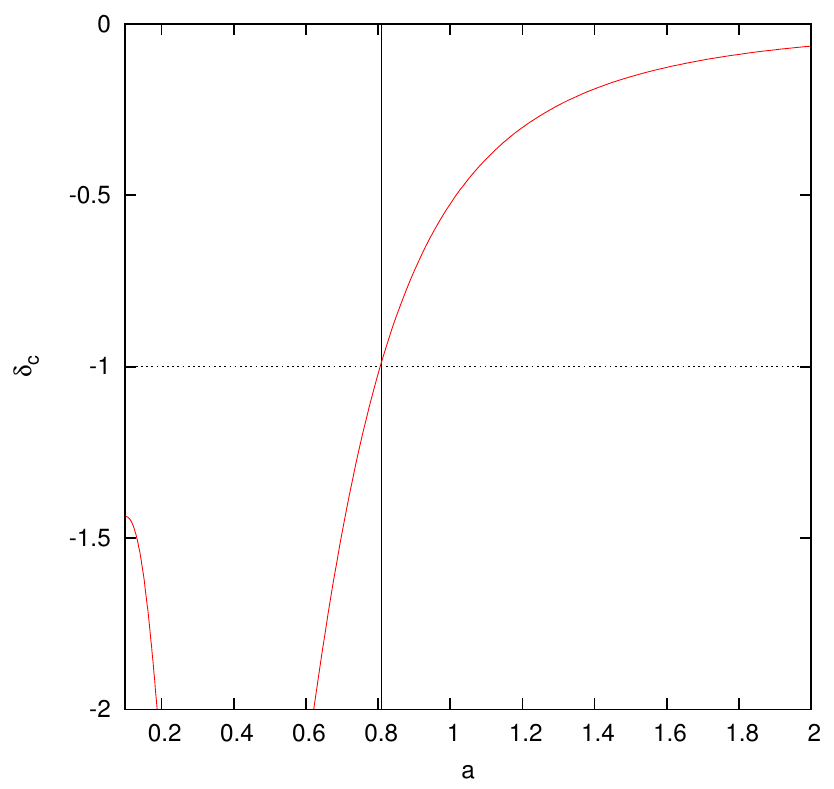}
\caption{The allowed density contrast $\delta_c$ of a void versus scale factor $a$ for which the condition Eq.~(\ref{breakdown_cond}) is satisfied for the Cubic Galileon model. For $a> 0.81$ we start to see $\delta_c > -1$ and in the far future ($a\to \infty$) we have $\delta_c \to 0$. At the present time, $a=1$, the analytic quasi-static field profile does not exist today for voids with density contrast $\delta \lesssim -0.52$.}
\label{fig:dencondition}
\end{figure}

For the Cubic Galileon model, however, the condition breaks down at late times in voids. For a void with density contrast $\delta_0$ the condition reads
\begin{align}\label{breakdown_cond}
\delta_0 \geq -\frac{9\beta_1\beta_2 a^3}{8\Omega_m}
\end{align}
In Fig.~(\ref{fig:dencondition}) we show the time evolution of this condition as a function of the scale-factor. At $a \gtrsim 0.8$ the condition starts to break down for the deepest voids $\delta_0 \simeq -1$ and at the present time the condition is violated for all voids with $\delta_0 < -0.5$. 

This was first noted in \cite{2013JCAP...10..027B} where the {\it N}-body simulations, based on solving Eq.~(\ref{full_kg_qs}), broke down in deep cosmic voids close to the present time. The authors choose to make an ad hoc fix to be able to integrate the equations until today with the argument that even though it breaks down in voids, whatever happens there should not significantly alter clustering statistics such as the halo mass function and the matter power-spectrum. It was also pointed out that this effect could to be due to the assumptions made, namely the quasi-static approximation.

We would like to know what happens to the solution when we enter this regime. Does it blow up or do the terms which were neglected when going to the quasi-static limit kick in and save the day? To answer such question we must move beyond the quasi-static limit and look at solutions to the equations when the time-evolution is taken properly into account. This will be the subject in the next section.

\section{Beyond The Quasi-Static Approximation}\label{sect:beyondqs}
In this section we will derive the full evolution equations, including metric perturbations, for our models. We are, for simplicity going to focus on spherically symmetric systems relevant for cosmology. The presentation in this section will focus on the Cubic Galileon. The non-linear terms in the Galileon models vanish for 1D configurations so the spherical symmetric case is the simplest setup for which we have the Vainshtein mechanism present. The equations and analysis for the DGP model is very similar so we will present it in Appendix \ref{app:dgp} instead. 

We work with the perturbed FRLW metric in the Newtonian Gauge 
\begin{align}
ds^2 = -dt^2(1+2\Psi) + a^2(t)(1-2\Phi)(dr^2 + r^2d\Omega^2)
\end{align}
To begin with we are going to make as few approximations as possible:
\begin{itemize}
\item The metric potentials $\Phi,\Psi$ are expanded to first order in perturbation theory.
\item We neglect all factors of $\Phi,\Psi$ that don't contain a derivative, i.e. $(1+\Phi) \simeq 1$. These terms only renormalize the terms we already have by a very small factor $\lesssim 1+\mathcal{O}(10^{-5})$, for the cosmological case, so it is safe to ignore them.
\item The scalar field $\phi = \overline{\phi} + \delta\phi$ is treated exactly, no perturbative expansion and no quasi-static approximation is applied.
\end{itemize}
In Appendix \ref{app:useful} we present expressions for the Christoffel-symbols, the Riemann tensor and derivative operators on the scalar field needed to derive the equations presented below.

\subsection{The dimensionless field equations}
Before we present the equations we will solve we will first introduce our dimensionless code-units. Our code variables are defined as
\begin{align}
x \equiv \log a,~~~y \equiv \frac{r}{R},~~~\omega \equiv  \frac{d\phi}{dy},~~~Q \equiv \frac{1}{H_0}\frac{d\phi}{dt}
\end{align}
where $R$ is the size of the simulation box at the present time and $\sigma \equiv \frac{1}{a(H_0 R)}$. Further we will also use $\nabla_y^2\phi = \frac{1}{y^2}\frac{d}{dy}\left(y^2\omega\right) = \frac{d^2\phi}{dy^2} + \frac{2}{y}\frac{d\phi}{dy}$ to simplify the notation. An overbar will denote a background quantity and $E \equiv \frac{H}{H_0}$ is the Hubble factor in units of the present value. We have also implemented physical coordinates $y_{\rm phys} = \frac{ar}{R} = ay$ and to get the equations in these coordinates the only change one has to make is  the replacement $\sigma \to a\sigma$ in all the equations below.

The first equation that is needed is the perturbed part of the Einstein equation, $\delta G^{(00)} = \delta T^{(00)}$. This gives us the Poisson equation for $\Psi$ which in code units can be written
\begin{widetext}
\begin{align}\label{poisson_eq_full}
\sigma^2\nabla_y^2\Psi = & \frac{3}{2}\Omega_ma^{-3}\delta_m + \frac{c_2}{4}(Q^2 - \overline{Q}^2 + \sigma^2\omega^2) + 3Ec_3(Q^3 - \overline{Q}^3) - c_3(Q^2 - \sigma^2\omega^2)\sigma^2\nabla_y^2\phi \nonumber\\
&- c_3EQ\sigma^2\omega^2 - \frac{2\sigma^4\omega^3c_3}{y} +\left[3E^2 - 3EQ^3c_3 + c_3EQ\sigma^2\omega^2\right]\left(\frac{d\Psi}{dx}\right) + c_3\omega\sigma^2\left[Q^2 +\sigma^2\omega^2\right] \left(\frac{d\Psi}{dy}\right)
\end{align}
\end{widetext}

For the anisotropic stress we  take $\delta G^r_r - \frac{1}{3}\delta G^i_i = \delta T^0_0 - \frac{1}{3}\delta T^i_i$ which gives us: 
\begin{widetext}
\begin{align}
\sigma^2\nabla_y^2(\Psi-\Phi) - \frac{3\sigma^2}{y}(\Psi-\Phi)' =  2c_3\sigma^2\omega^2\left[3EQ + \sigma^2\nabla_y^2\phi - \frac{4\sigma^2\omega}{y} + \frac{c_2}{2c_3} + E\left(\frac{dQ}{dx}\right)\right]- 4c_3\sigma^2\omega Q\left(\frac{dQ}{dy}\right)\nonumber\\ + 2c_3\sigma^2\omega\left[ - EQ\omega\left(\frac{d\Phi}{dx}\right) + (2Q^2-\sigma^2\omega^2)\left(\frac{d\Phi}{dy}\right) - 3EQ\omega \left(\frac{d\Psi}{dx}\right) + 3\sigma^2\omega^2\left(\frac{d\Psi}{dy}\right)\right]
\end{align}
\end{widetext}
The right hand side vanishes if $\omega = 0$ and, in that case, we expect $\Phi \simeq \Psi$ to hold to a high accuracy.  Finally we have the Klein Gordon equation for $\phi$ which can be written as an evolution equation for $Q = \frac{1}{H_0}\frac{d\phi}{dt}$:
\begin{widetext}
\begin{align}\label{eq:eom_code_units}
&E\left[12EQ + \frac{c_2}{c_3} - 4\sigma^2\nabla_y^2\phi - 12 EQ \left(\frac{d\Psi}{dx}\right) + 4\sigma^2\omega\left(\frac{d\Psi}{dy}\right)\right]\left(\frac{dQ}{dx}\right) = \nonumber\\
& - \frac{3EQc_2}{c_3} + 8E\sigma^2\omega\left(\frac{dQ}{dy}\right) - 4\sigma^2\left(\frac{dQ}{dy}\right)^2 + \sigma^2\nabla_y^2\phi\left[\frac{c_2}{c_3} + 8EQ - \frac{8\sigma^2\omega}{y}\right]-6\left(\frac{dE}{dx}\right) E Q^2\nonumber\\
 &- 18E^2 Q^2 + 2\left(\frac{dE}{dx}\right)E\sigma^2\omega^2 + 2E^2\sigma^2\omega^2 + \frac{12\sigma^4\omega^2}{y^2} - 2\sigma^2\nabla_y^2(\Psi-\Phi) + 2\sigma^2\nabla_y^2\Psi(Q^2 + \sigma^2\omega^2)\nonumber\\
 &+ E\mathcal{A}_1\left(\frac{d\Phi}{dx}\right)+ E\mathcal{A}_2\left(\frac{d\Psi}{dx}\right)+ \sigma\mathcal{A}_3\left(\frac{d\Phi}{dy}\right) +\sigma\mathcal{A}_4\left(\frac{d\Psi}{dy}\right) + \sigma E\mathcal{A}_5\left(\frac{d^2\Psi}{dydx}\right) + E^2\mathcal{A}_6\left(\frac{d^2\Psi}{dx^2}\right)
\end{align}
where $\mathcal{A}_{1-6}$ are functions given in Appendix \ref{app:aterm}.
\end{widetext}

Removing the perturbations $\omega,\Phi\to 0$, $Q \to \overline{Q}$ we get the homogenous scalar field equation for the background
\begin{widetext}
\begin{align}
\frac{d\overline{Q}}{dx} = \left\{ -\frac{3\overline{Q}Ec_2}{c_3} - 18\overline{Q}^2E^2 - 6\overline{Q}^2E\frac{dE}{dx}\right\}\times \left(\left(\frac{c_2}{c_3} +  12E\overline{Q}\right)E\right)^{-1}
\end{align}
which can be written using Eq.~(\ref{hubbeleqs}) as
\begin{align}
\frac{d\overline{Q}}{dx} = \left\{ -\frac{3\overline{Q}Ec_2}{c_3} - 12\overline{Q}^2E^2 +3\Omega_m e^{-3x} \overline{Q}^2 + 2c_2\overline{Q}^4 + 6c_3\overline{Q}^5E\right\}\times \left(\left(\frac{c_2}{c_3} +  12E\overline{Q} + 6c_3\overline{Q}^4\right)E\right)^{-1}
\end{align}
\end{widetext}
This equation can be seen to be identical to Eq.~(\ref{eombackground}) and serves as a simple consistency check. Lastly we need the Hubble equation Eq.~(\ref{hubbeleqs}) on a dimensionless form:
\begin{align}
E &= c_3 \overline{Q}^3 + \sqrt{c_3^2\overline{Q}^6 + \frac{c_2}{6}\overline{Q}^2 + \Omega_m e^{-3x}}
\end{align}
The equations above form a closed system, as long as $\delta_m(y,x)$ is given, and are all we need for our numerical implementation.

\subsection{Conserved Charge}

The Galileon action is invariant under the symmetry $\phi \to \phi + c + d_{\mu}x^{\mu}$ and consequently, through Noether's theorem, there is a conserved current $J^{\mu}$ in terms of which which the equation of motion can be written 
\begin{align}
\nabla_{\mu}J^{\mu} = \frac{d}{dt}(\sqrt{-g}J^0) + \frac{d}{dr}(\sqrt{-g}J^r) = 0
\end{align}
where the first equality only holds for spherical symmetry. Associated with this current is a conserved change density\footnote{The integrals here are to be interpreted as first taken over a fixed volume in co-moving coordinates and then taking the limit of the volume going to infinity. For numerical simulations with periodic boundary conditions the integrals are simply taken over the simulation volume.}
\begin{align}
\rho_{\rm Noether} \equiv \frac{\int \sqrt{-g}J^0dr}{\int r^2 dr} 
\end{align}
where
\begin{align}
J^{\mu} = \nabla^{\mu}\phi\left(\frac{c_2}{c_3} + \frac{2}{M^3}\square\phi\right) - \frac{1}{M^3}\nabla^{\mu}(\nabla\phi)^2
\end{align}
and in particular
\begin{align}
\sqrt{-g}J^{0} = e^{x}\left(\frac{y}{\sigma}\right)^2\left[Q\left(\frac{c_2}{c_3} + 6EQ - 2\sigma^2\nabla_y^2\phi\right) \right.\nonumber\\
\left.+ 2H\frac{d\Phi}{dx}(\sigma^2\omega^2 - 3Q^2) - 2E\sigma^2\omega^2 + 2\sigma^2\omega\frac{dQ}{dy}\nonumber\right.\\
\left. +2Q\sigma^2\omega\frac{d}{dy}(\Phi-\Psi)\right]
\end{align}
For the background the conserved charge density becomes
\begin{align}
\overline{\rho}_{\rm Noether} = e^{3x}\left(\frac{c_2}{c_3} + 6E\overline{Q}\right)\overline{Q}
\end{align}
By taking the derivative of this equation, $\frac{d\overline{\rho}_{\rm Noether}}{dx} = 0$, we recover the background equation of motion Eq.~(\ref{eombackground}). We can use the conservation of the Noether charge density as a test of our code by monitoring the constancy of
\begin{align}\label{noether_eps}
\epsilon \equiv \frac{\rho_{\rm Noether} - \overline{\rho}_{\rm Noether}}{\overline{\rho}_{\rm Noether}}
\end{align}

\subsection{What terms are expected to be small?}\label{sect:smallterms}
One of the aims with this paper is to rigorously classify what are terms in the (long) equations above that can be safely neglected. We will first go through the standard arguments for why some terms should be small, then we will study this numerically in the next section.

All the terms in $\mathcal{A}_{1,2,6}$ can be found elsewhere in the equation of motion so these terms are expected to be negligible whenever $\left|\frac{d\Phi}{dx}\right|,\left|\frac{d\Psi}{dx}\right|, \left|\frac{d^2\Psi}{dx^2}\right| \ll 1$. In a cosmological context the potentials $Y = \Phi,\Psi$ usually evolve over a Hubble time which makes $\left|\frac{dY}{dx}\right| \sim |Y| \ll 1$. The terms containing derivatives of the gravitational potential need a closer inspection. If we define $\left<\delta_m(r)\right>$ to be the average matter density within a radius $r$ we roughly have
\begin{align}
\frac{d\Psi}{dr} \sim \frac{a^{-1}}{2}\Omega_m\left<\delta_m(r)\right> r
\end{align}
and likewise for $\Phi$. This leads to the constraint
\begin{align}
r \ll \frac{a}{\Omega_m} \frac{6000}{\left<\delta_m(r)\right>} \text{ Mpc}/h
\end{align}
which is satisfied for most cosmological applications. We also expect perturbations in the scalar field to be small. In particular we expect
\begin{align}
\left|\delta\phi\right| \ll 1,~~~~~~\left|\frac{d\phi}{dr}\right| \ll \left|\frac{d\overline{\phi}}{dt}\right|,~~~~~~\left|\dot{\delta\phi}\right| \ll \left|\dot{\overline{\phi}}\right|
\end{align}
If the first condition is violated then we are outside the realm of the effective theory which we are working with. If the second or third condition is violated it means that the clustering of the scalar field will back-react and thereby changing the background evolution of the Universe.

Lastly we have the anisotropic stress. If there is no perturbations in the scalar field this vanishes identically and in general we have
\begin{align}
\nabla_y^2(\Phi - \Psi) \sim \mathcal{O}(\omega^2Q)
\end{align}
Given that $\omega^2 \sim \Psi^2$ and $EQ = \mathcal{O}(1)$ we also expect the anisotropic stress to be negligible.

We finish by pointing out that there is one term neglected in the quasi-static limit that could potentially be large and that is the cross term $\frac{d^2\phi}{dtdr}$. We would generally expect $\frac{d^2\phi}{dtdr} \sim H \frac{d\phi}{dr}$ which is comparable with other terms in the equation of motion.

\section{Numerical Implementation}\label{sect:numberical}
The fundamental variables for the scalar field in our code are the 'position' $\phi$ and 'velocity' $Q = \frac{1}{H_0}\dot{\phi}$ which are discretised on a linearly spaced grid going from $y=0$ to $y=1$ with $N$ grid cells. 

\subsection{Leap-Frog Integrator}
Our first method to solve the system is to propagate the variables $\{\phi,Q\}$ in time using a staggered Leap Frog algorithm. Starting with $Q_{n-1/2}$ and $\phi_{n}$ we first propagate $Q$ one step using
\begin{align}
Q_{n+1/2} = Q_{n-1/2} + \left(\frac{dQ}{dx}\right)_{n} \Delta x
\end{align}
and then use the result to propagate $\phi$ using
\begin{align}
\phi_{n+1} = \phi_{n} + [Q E]_{n+1/2} \Delta x
\end{align}
From $\phi_{n+1}$ we calculate $\omega_{n+1} = \frac{d\phi_{n+1}}{dy}$ and other spatial derivatives depending on $\phi$ using a five-point stencil which is fourth order accurate\footnote{We have also tried using fifth-order splines, but this is much slower and was not found to significantly improve the accuracy of the solution.}. This is then again used in the next time-step to evaluate $\frac{dQ_{n+1}}{dx}$. 

\subsection{Newton Gauss Seidel Integrator}
The Klein Gordon equation Eq.~(\ref{eq:eom_code_units}) and Eq.~(\ref{eq:dgpfull}) can be written on the schematic form $\mathcal{L} = A(\phi,\dot{\phi},\phi',\phi'',t)\ddot{\phi} - B(\phi,\dot{\phi},\dot{\phi}',\phi',\phi'',t) = 0$. Discretizing the operator $\mathcal{L}$ on our grid it becomes a non-linear equation for $\phi(y_i, t_n)$. Our second method is to solve the equation
\begin{align}
\mathcal{L}(\phi(y_i, t_n)) = 0
\end{align}
at each time-step using Newton-Gauss-Seidel relaxation
\begin{align}
\phi^{\text{new}}(y_i, t_n) = \phi^{\text{old}}(y_i, t_n) - \frac{\mathcal{L}(y_i, t_n)}{\partial\mathcal{L}(y_i, t_n)/\partial \phi(y_i, t_n)}
\end{align}
with red-black colouring of the grid nodes. Note that this method is {\it much} slower than our main method (the leap-frog) as we must solve an algebraic equation at every step as opposed to simply updating a value.

Time-derivatives, such as $\dot{\phi}$ and $\ddot{\phi}$, are calculated using the backward-scheme
\begin{align}
\dot{\phi}(t_n) &= \frac{\phi(t_n) - \phi(t_{n-1})}{\Delta t}\\
\ddot{\phi}(t_n) &= \frac{\phi(t_n) - 2\phi(t_{n-1}) + \phi(t_{n-1})}{(\Delta t)^2}
\end{align}
for second order accuracy. Grid derivatives are calculated as 
\begin{align}
\phi'(y_i) &= \frac{\phi(y_{i+1}) - \phi(y_{i-1})}{2\Delta y}\\
\phi''(y_i) &= \frac{\phi(y_{i+1})  - 2\phi(y_{i}) + \phi(y_{i-1})}{(\Delta y)^2}
\end{align}
for second order accuracy. We have also implemented and tested higher order discretizations, but our second order implementation was found to be well behaved and no visible difference in the results was found when we included higher order terms.

\subsection{Time-steps and initial conditions}
The time-step $\Delta x$ is initially fixed, but can be adjusted adaptively depending on how close the denominator $B$ in the equation for $\frac{dQ}{dx} = \frac{A}{B}$ is to zero: we cut the log-time-step proportional to the value of the denominator. This only applies for the void simulations of the Cubic Galileon. In all other cases we used a fixed $\Delta x$.  The initial values we choose depend on the application, but in general we just put $\phi = 0$ and $Q = \overline{Q}$ at the initial time-step.

Most of our numerical solutions have been derived using both of the integration methods to cross-check the results. In general the NGS solver was found to be the most stable one; allowing larger time-steps than the leap-frog without breaking down, but then again it is much much slower.

\subsection{Treatment of the matter sector}

The evolution (gravitational collapse) of matter will in general be affected by both the standard Newtonian force and the fifth-force from the scalar field\footnote{Matter moves under the force given by the potential $\Psi$, but in the non-relativistic weak-field limit we can separate the potential $\Psi$ into the standard Newtonian part $\Psi_N$ and a fifth-force $\propto \nabla\phi$.}. Both of these terms should be calculated simultaneously. However, since the purpose of this paper is to study the evolution of the scalar field we have simply chosen to impose a density profile and then study the evolution of the scalar field given this profile. This simplifies the numerical solution as we don't have to evolve the matter sector. This is a self-consistent procedure as one can easily imagine (in principle) that the profile is being set up and held together by non-gravitational forces. We choose the mexican-hat profile
\begin{align}
\delta_m(a,r) = \delta_0\left(1 - \frac{r^2}{3\sigma^2}\right)e^{-\frac{r^2}{2\sigma^2}}f(a)
\end{align}
where $f(a)$ is some function regulating the amplitude of the matter perturbation as a function of time. This profile has the advantage that $\int_0^\infty \delta_m(t,r) 4\pi r^2 dr = 0$ so our simulations will not have any excess matter compared to the cosmic mean (which could possibly bias the results). Unless stated otherwise the main choice for $f$ is $f(a) = a^3$ for time-evolving profiles and $f(a) = 1$ for static profiles.

\subsection{Boundary conditions}

The Lagrangian for our models does not have any mass-term which means that, in the linear limit, the scalar field will have waves that are only weakly damped (by friction terms that depend on the profile we have at any moment) as the scalar waves propagate. This can cause a problem when waves are created in the box and start propagating out towards the boundary of the simulation-box. If we employ standard boundary conditions $\frac{d\phi}{dr} = 0$ at $r=R$ the outgoing waves will reflect at the boundary and start traveling inwards.

\begin{figure}
\includegraphics[width=0.9\columnwidth]{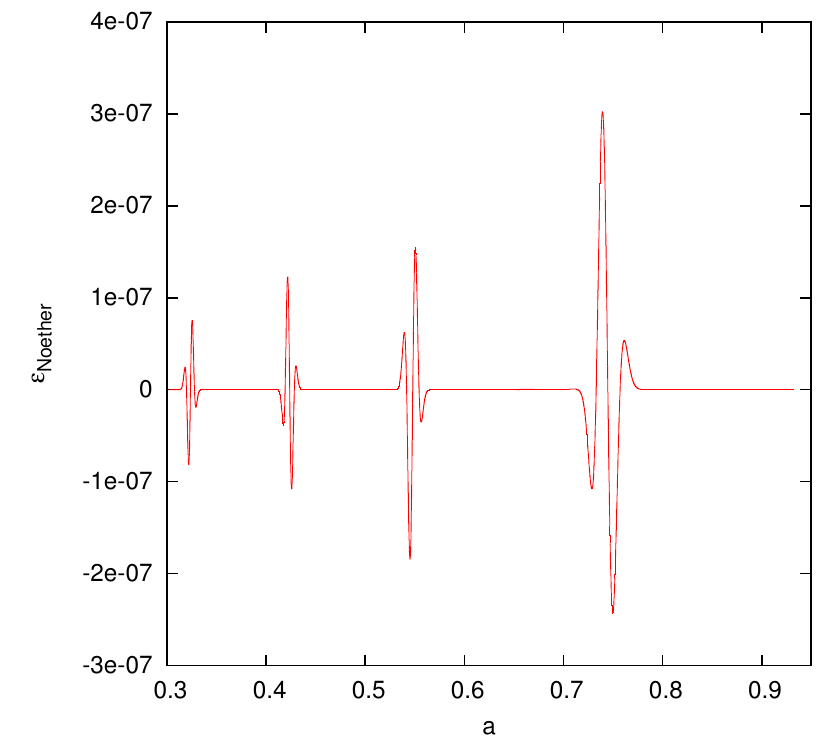}
\caption{Evolution of the Noether charge, Eq.~(\ref{noether_eps}), for a 'cluster' simulation. The spikes happen when outgoing waves reflect of the boundary in our simulation box. However, after the reflection the charge returns back to the value where it started beforehand.}
\label{fig:noether}
\end{figure}

The reflection of these can be a potential problem and there are a few ways to mitigate it. The simplest brute-force solution is to make the box so large that the waves won't have time to reflect back in on the time scale of the problem we are solving. A more sophisticated solution is to try to construct non-reflecting boundary conditions or even introduce artificial damping terms into the equations of motion. We have opted for the first, and simplest, approach and have investigated its validity by 
using the conservation of the Noether current Eq.~(\ref{noether_eps}) during the evolution. In Fig.~(\ref{fig:noether}) we show the conserved charge for a test-case where a mexican-hat  profile (see the previous section) with $\delta_0 = 100$. As the outgoing waves reflect from the boundary the Noether charge is seen to change, but returns to its previous (small) value after the reflection has ended and the wave is traveling back in again. We have also explicitly checked, by running simulations using a very large box so that outgoing waves never get back in again. We have found that the main results of this paper do not depend critically on this choice of boundary condition.

\section{Code Tests}\label{sec:codetest}
Here we present some of the tests we have performed to ensure that our code is working correctly.

\subsection{No perturbations}
\begin{figure}
\includegraphics[width=0.9\columnwidth]{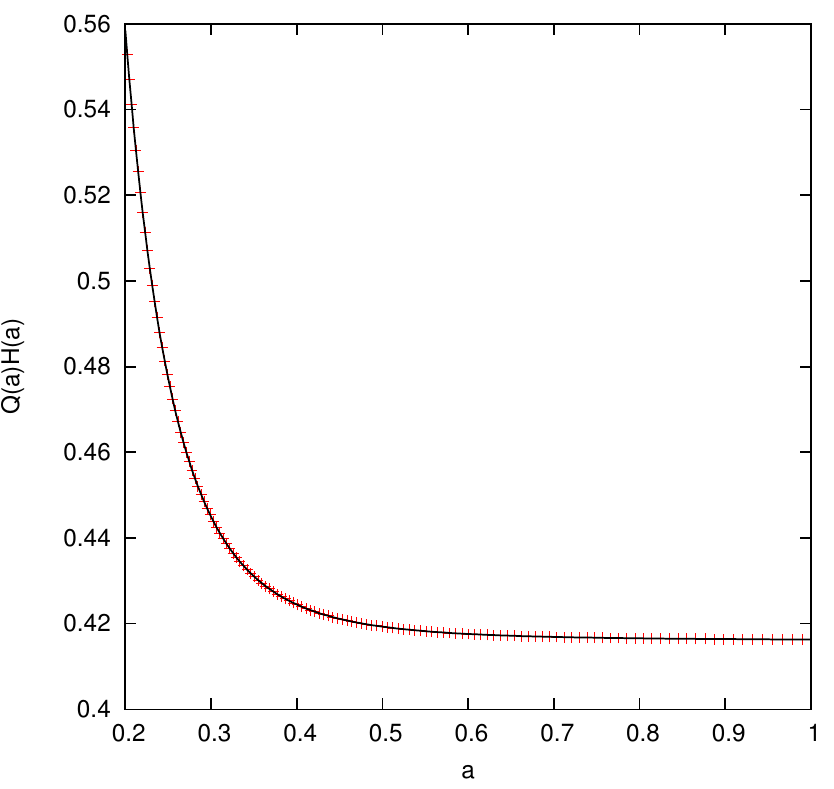}
\caption{The solution for $EQ = \frac{H(a)}{H_0^2} \frac{d\phi}{dt}$ in a simulation where $\delta_m = 0$. This effectively means that we are solving the background equation of motion in each grid-cell independently. We plot the average value $\left<EQ\right>$ (symbols) together with the analytical result (solid line).}
\label{fig:nopert}
\end{figure}

The first test we perform is to put $\delta_m \equiv 0$. This effectively solves the background equation of motion in every grid cell and $Q = \overline{Q}$ is the expected result. The result displayed in Fig.~(\ref{fig:nopert}) shows perfect agreement with expectations.

\subsection{A small perturbation}
Building on the previous test we introduce a small perturbation in the field and study the propagation of the waves it creates. Linearizing the equation of motion around the cosmological background we find 
\begin{widetext}
\begin{align}\label{lineareom}
\ddot{\delta\phi}\left( \frac{c_2}{c_3} + 12E \overline{Q} \right) - \sigma^2\nabla_y^2\delta\phi \left(\frac{c_2}{c_3} + 8 E\overline{Q} -2c_3\overline{Q}^4 + 4H \frac{d\overline{Q}}{dx}\right)\nonumber\\
+ \dot{\delta \phi}\left(12 E^2 \frac{d\overline{Q}}{dx} + \frac{3Ec_2}{c_3} + 36\overline{Q} E^2 + 6\overline{Q}\frac{dE^2}{dx} - c_2\overline{Q}^3 - 18Ec_3\overline{Q}^4\right) = 3\overline{Q}^2a^{-3}\Omega_m\delta_m
\end{align}
\end{widetext}
The corresponding equation(s) for DGP can be found in Appendix \ref{app:dgp}. From this we can read of the speed of sound, in physical coordinates, as
\begin{align}
c^2 = \frac{\frac{c_2}{c_3} + 8 E\overline{Q} -2c_3\overline{Q}^4 + 4E \frac{d\overline{Q}}{dx}}{ \frac{c_2}{c_3} + 12E \overline{Q}}
\end{align}
for the Cubic Galileon. We test the code by taking $\delta_m = 0$, placing the system in the cosmological background attractor $Q = \overline{Q}$ and then add a small, one-wavelength, perturbation $\delta\phi \propto \sin(2\pi \frac{y-y_*}{dy})\theta(y-y_*)$, where $\theta$ is the Heaviside function.

Solving the wave equation analytically is not easy as both the speed of sound and the friction term depends on time. However, we can derive an analytical approximation,
\begin{align}\label{waveapprox}
\delta\phi  \sim \frac{e^{-\frac{1}{2}\int \frac{D(x)}{E(x)}dx}}{r} \sin\left(\frac{2\pi}{dy}(y-y_* \pm ct(x))\right)\nonumber\\
\times \theta(y-y_* \pm ct(x))\theta(y_*+dy \mp ct(x) -y)
\end{align}
where $ct(x) = \int \frac{c(x)}{E(x)}dx$ and $D(x)$ is the term in front of $\dot{\delta\phi}$ divided by the term in front of $\ddot{\delta\phi}$.

\begin{figure*}
\includegraphics[width=1.0\columnwidth]{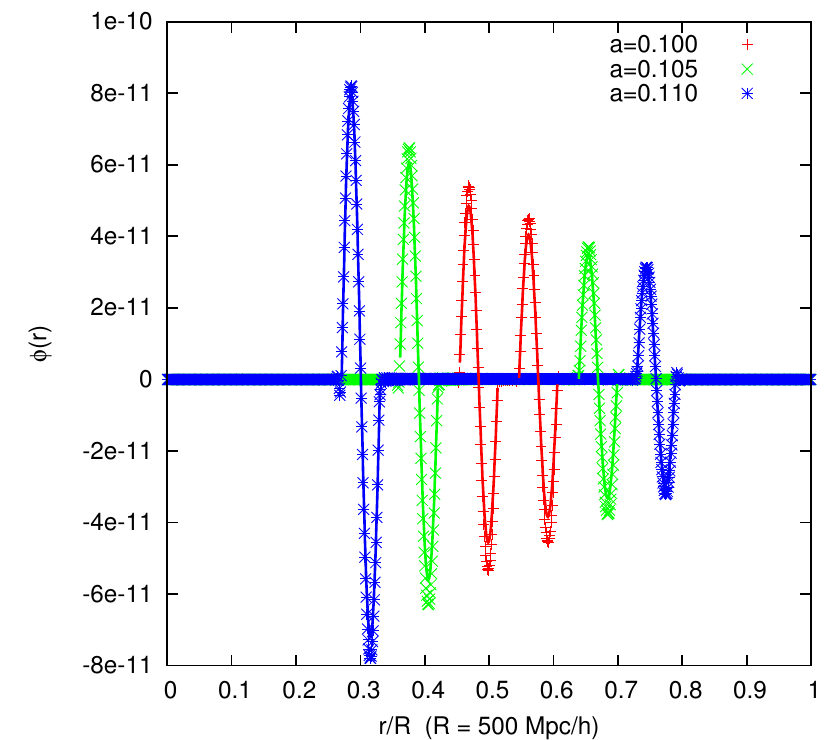}
\caption{The evolution of a one-wavelength sine wave perturbation $\delta\phi$, starting from $r/R=0.5$. We show both the ingoing and outgoing wave, at three snapshots (in the order blue, green and then red) after the initial release together with our analytical approximation Eq.~(\ref{waveapprox}) for the self-accelerating DGP model with $r_cH_0 = 1.35$. The difference between the full non-linear solution and the solution found by implementing the linearized evolution equation is indistinguishable with both of our methods of integration.}
\label{fig:wave}
\end{figure*}

Since the approximation above is not exact we don't expect perfect agreement. We therefore also choose to implement and solve the linear field equation Eq.~(\ref{lineareom}) and use this to compare our result with (using both of our integration methods). In Fig.~(\ref{fig:wave}) we show the result of this test. The full equations agrees perfectly with the linearized solution and also fairly well with our analytical approximation.

\subsection{Convergence of the solution}

We have performed several tests to check how sensitive our results are to the grid size, the size of the time-steps and the integration method used in the calculation. The results shown below are simulations of a growing mexican hat void profile ($f(a) = a^3$) with final density contrast $\delta_0 = -0.5$.

\begin{figure*}
\includegraphics[width=0.75\columnwidth]{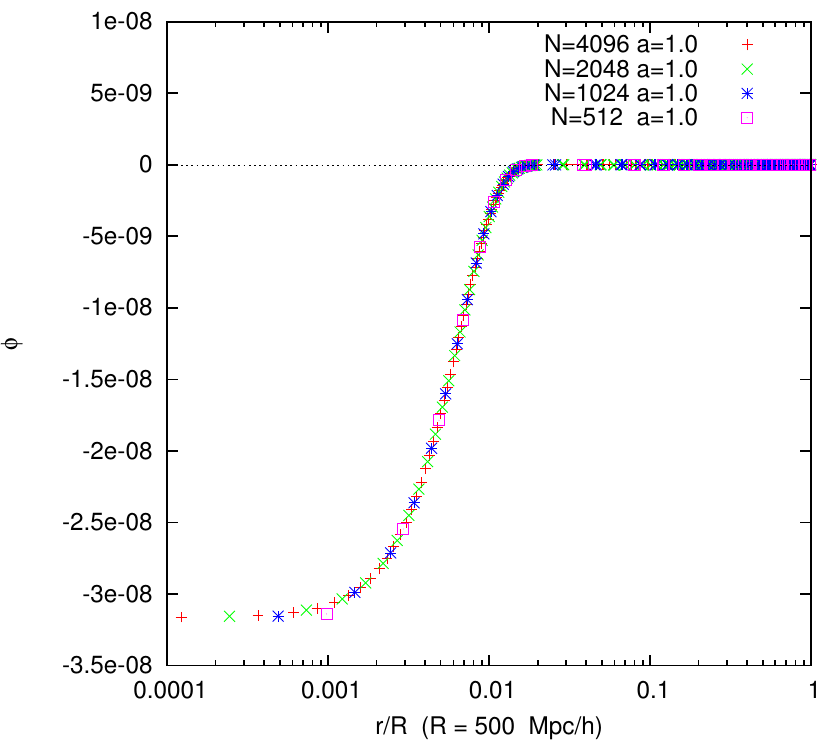}
\includegraphics[width=0.75\columnwidth]{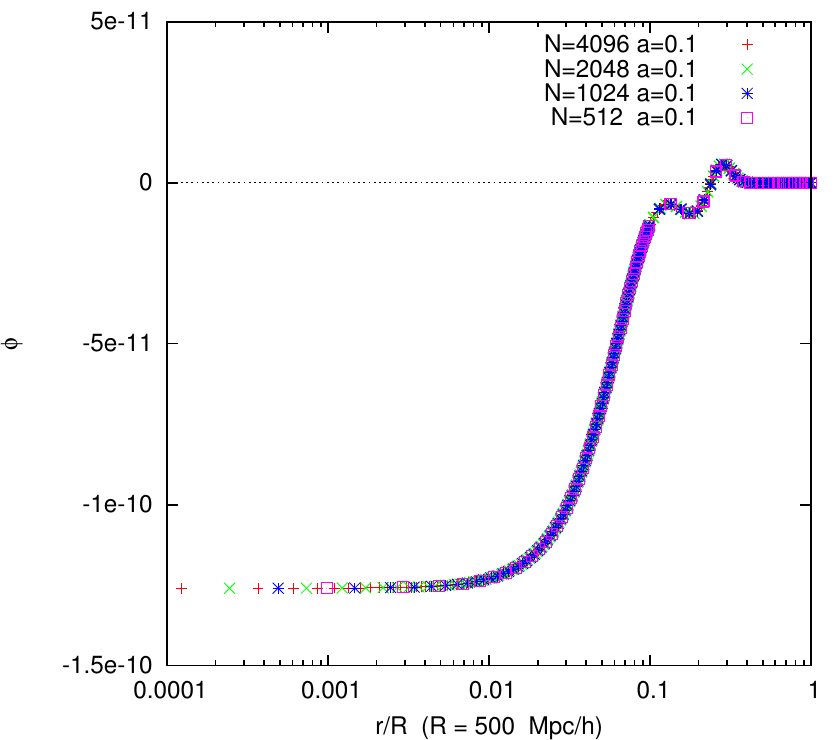}
\includegraphics[width=0.75\columnwidth]{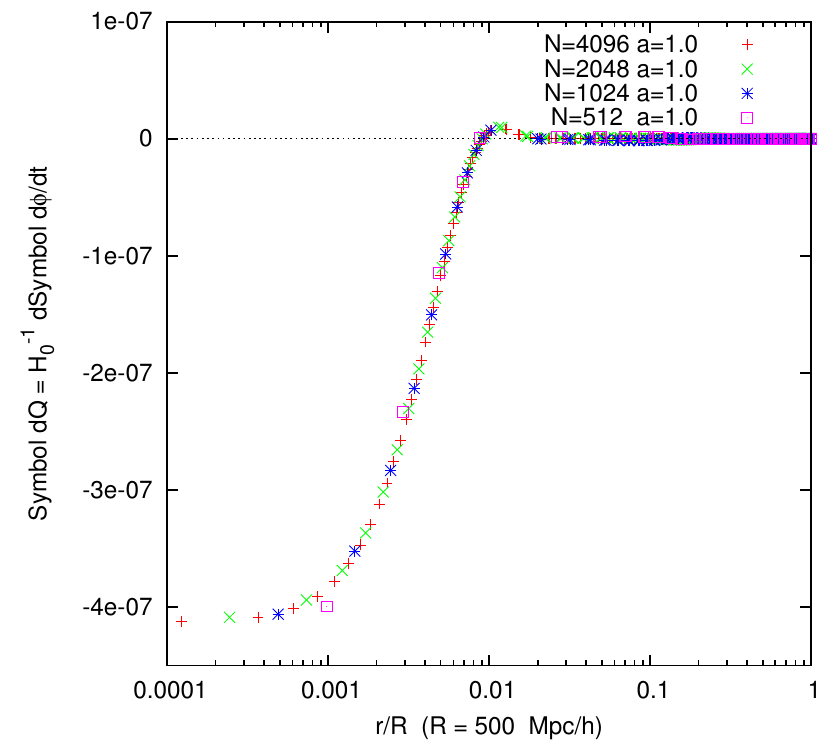}
\includegraphics[width=0.75\columnwidth]{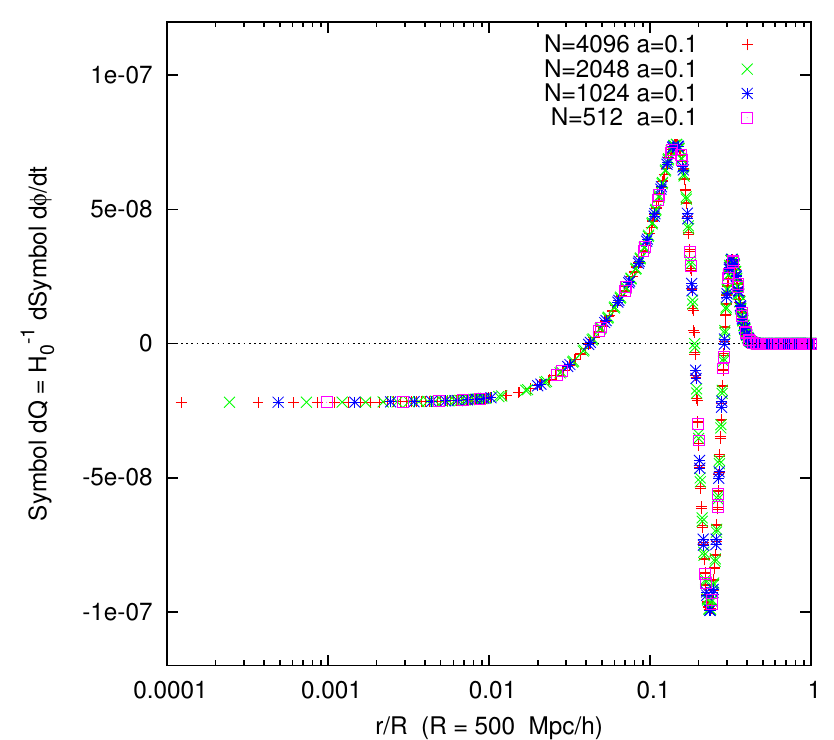}
\includegraphics[width=0.75\columnwidth]{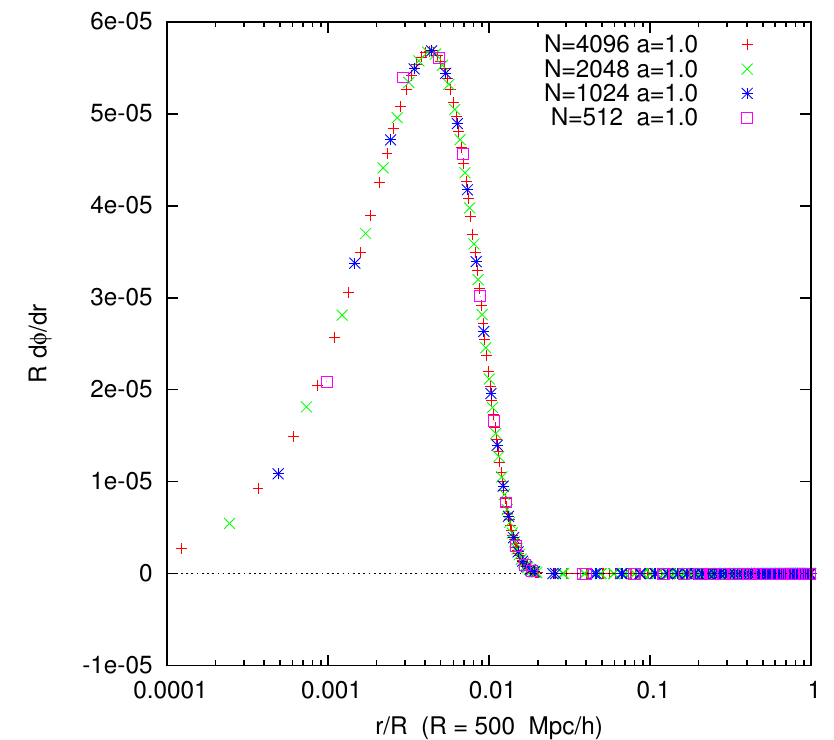}
\includegraphics[width=0.75\columnwidth]{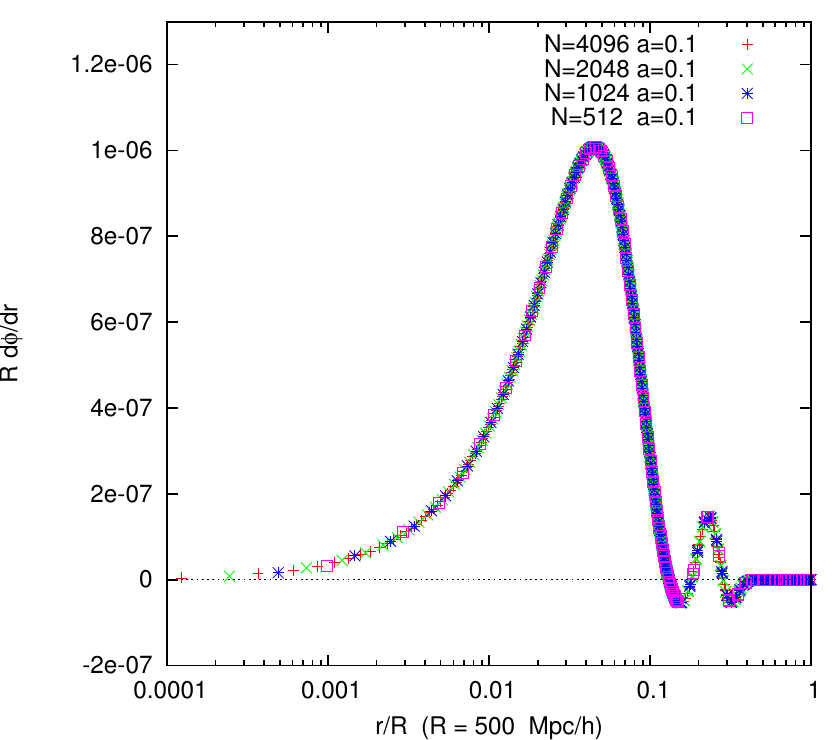}
\caption{Convergence test with respect to the number of grid nodes we use. Here we show the profiles $\phi(y)$ (top), $\delta Q = H_0^{-1}\frac{d\delta\phi}{dt}$ (middle) and $\omega = R \frac{d\phi}{dr}$ at $a=1.0$ (left) and $a=0.1$ (left) for four simulations using $N=512,1024,2048$ and $4096$ grid nodes respectively. The results agree very well at both times and all radii, with only some small differences visible for the $N=512$ case compared to the more refined simulations. These results are for the Cubic Galileon, but we get the same behavior for the DGP model.}
\label{fig:convtestgrid}
\end{figure*}

In Fig.~(\ref{fig:convtestgrid}) we show results from calculations using $N=512,1024,2048$ and $4096$ grid nodes respectively. The results agree very well across the different resolutions. The only real difference we found is in the first few time-steps where scalar waves are emitted (since we start off with $\phi = 0$) going out from the object at early times. The waves are created at $r/R =0$ and the coarser the grid the further out the wave will start (the minimum $r/R$ we can represent in the code is $(r/R)_{\rm min} = \frac{1}{N}$) thus giving it a head-start when comparing it with the more refined simulations.

\begin{figure*}
\includegraphics[width=0.75\columnwidth]{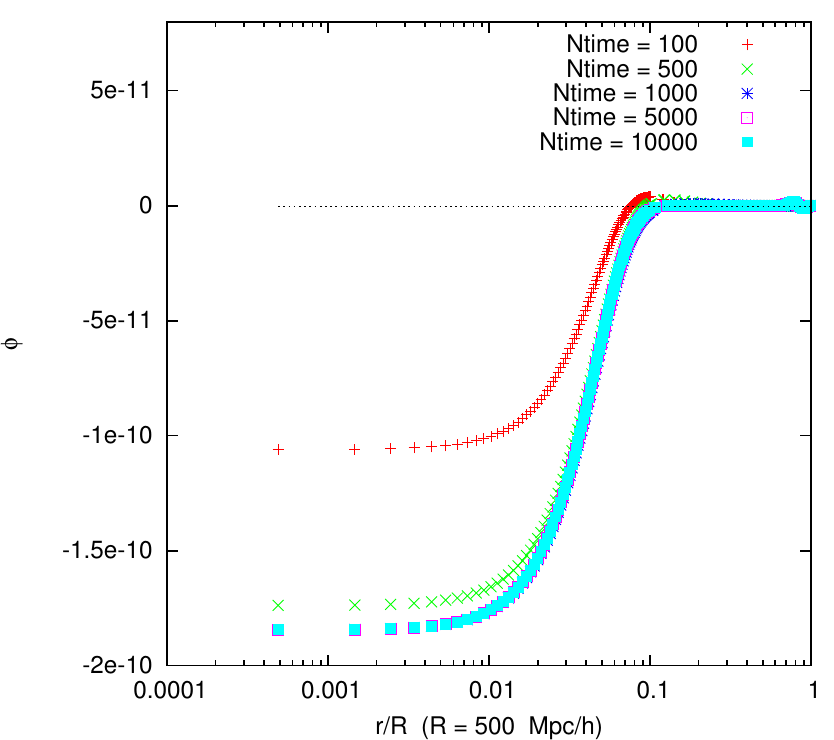}
\includegraphics[width=0.75\columnwidth]{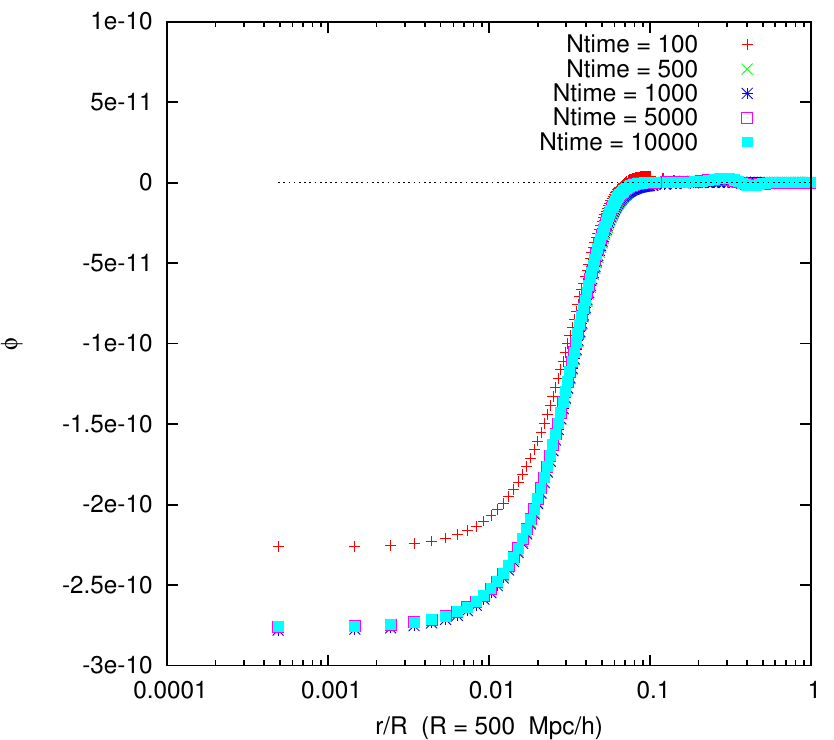}
\includegraphics[width=0.75\columnwidth]{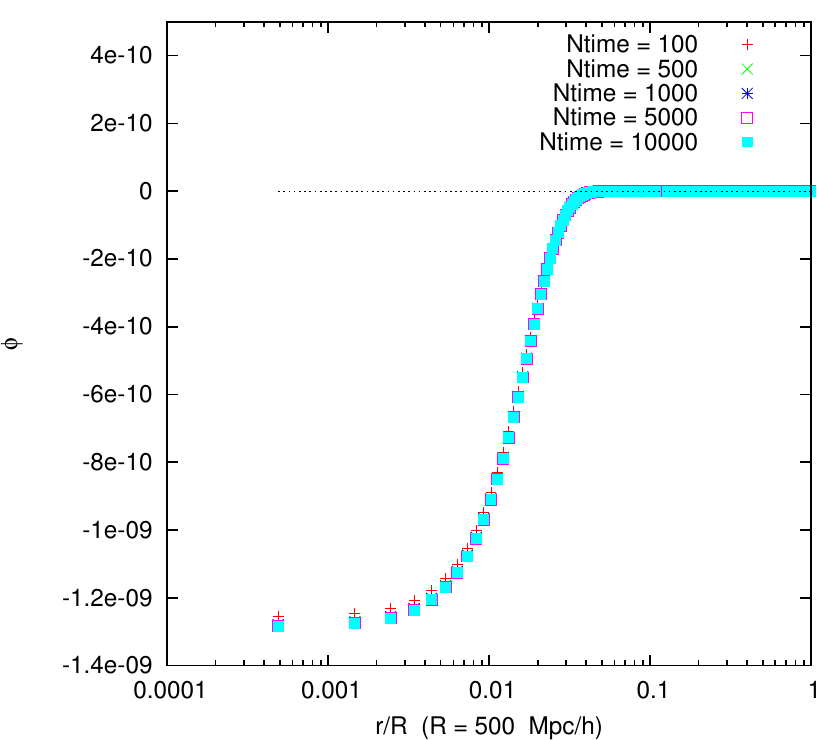}
\caption{Convergence test with respect to the number of time-steps we use. Here we show the field profile $\phi(r)$ at $a=0.15,0.2$ and $a=0.4$ for five different simulations using $N_{\rm time} = 100,500,1000,5000$ and $10000$ time-steps between $a=0.1$ and $a=1.0$. At early times, when the field is still evolving towards the quasi-static solution, the error can be quite large if we use too few time-steps. However, at later times when the field have settled close to the (evolving) quasi-static solution this difference has largely been washed away. These results are for the Cubic Galileon, but we get the same behavior for the DGP model.}
\label{fig:convtesttimes}
\end{figure*}

Next we have tested how the results depend on the size of the time-steps used in the computation. First of all we should mention that we are required to take very small time-steps in order for the solver not to break down. The leapfrog solver is more sensitive to the size of the time-steps than the NGS solver. The tests shown here are for the NGS integrator which allows us to go all the way down to $N=100$ time steps between $a=0.1$ and $a=1.0$. The results are shown in Fig.~(\ref{fig:convtesttimes}). With very few time-steps we see a difference in the solution compared to the simulations with the most time-steps. This difference seems to go away with time and the reason for this is that the quasi-static solution is an attractor for the system, however we are unable to resolve the oscillations of the field if we keep the number of time-steps small (small here means less than $1000$).  As we increase the number of time-steps the solution we find stops changing (including the oscillations we see in the field). For this particular test we found that roughly $10^5$ time-steps is needed to accurately track the evolution of the field at all times from $a=0.1$ till $a=1.0$.

\begin{figure}
\includegraphics[width=1\columnwidth]{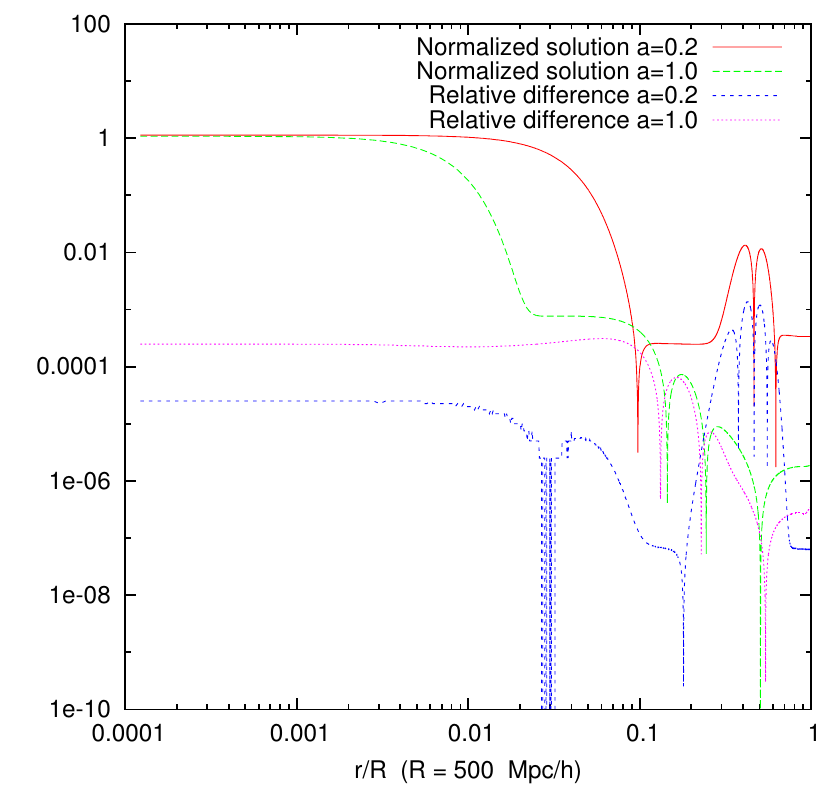}
\caption{The solution for the Cubic Galileon $\phi(r)$ normalized such that $\phi(0) = 1$ together with the difference between the solutions found by using the two different integration methods (leapfrog and NGS). For this test we used a $N=4096$ grid with $N_{\rm time} = 5\cdot 10^4$ time-steps from $a=0.1$ till $a=1.0$ and where the density profile was a growing void with $\delta_0 = -0.5$. We see that the difference stays below the $\sim 10^{-4}$ level whenever $\phi$ itself is larger than $\sim 10^{-4}$ times its maximum value.}
\label{fig:error}
\end{figure}

Finally we tested the difference between the two different integration methods keeping the time-steps equal. In Fig.~(\ref{fig:error}) we show the difference in the solution for $\phi(r)$ between the leapfrog and the NGS solver for simulations of a growing void. The agreement is excellent where it should be. It is only in regions where $\phi \approx 0$ that we see a small difference.

\section{Simulation Results}\label{sect:results}

In this section we will present the results of our simulations and try to answer the questions we posed in the introduction.

\subsection{Does the Vainshtain solution always forms?}
To try to answer the question in the title we impose an analytic density profile $\delta_m(x,y)$ and release the scalar field at $\phi = 0$ with $Q = \overline{Q}$ and follow the subsequent evolution.

\begin{figure*}
\includegraphics[width=0.75\columnwidth]{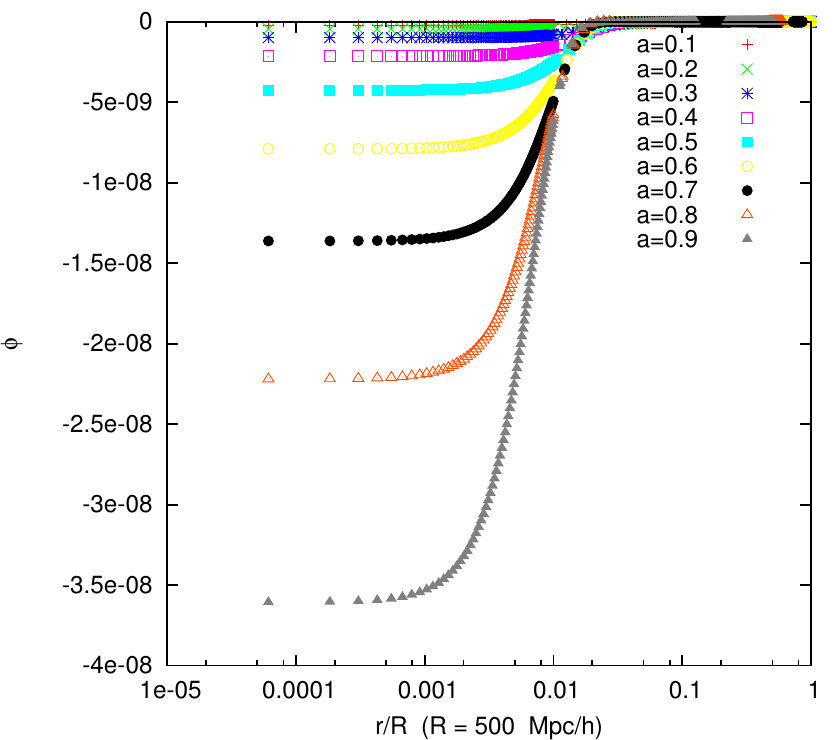}
\includegraphics[width=0.75\columnwidth]{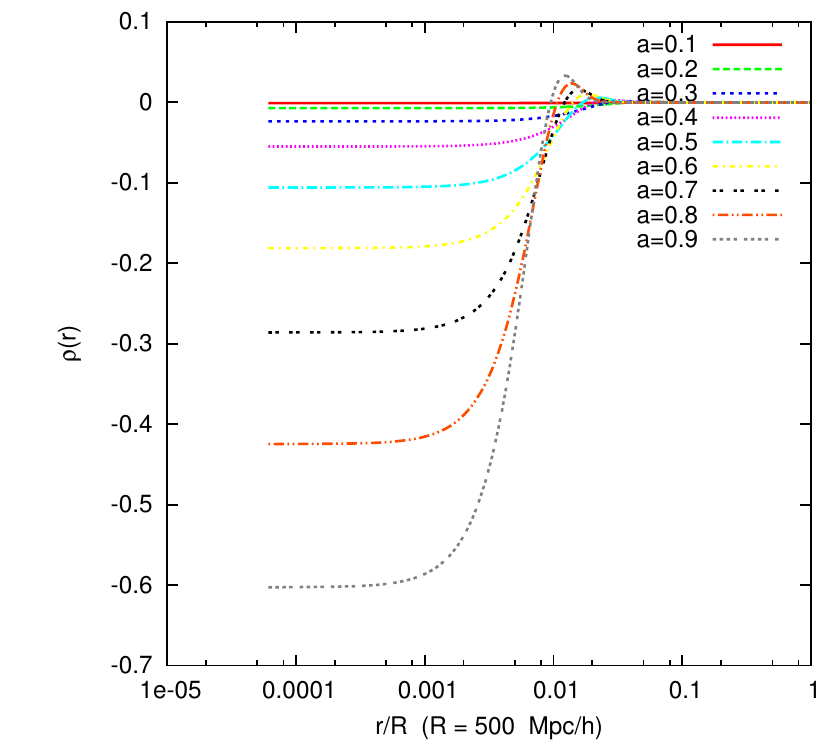}\\
\includegraphics[width=0.75\columnwidth]{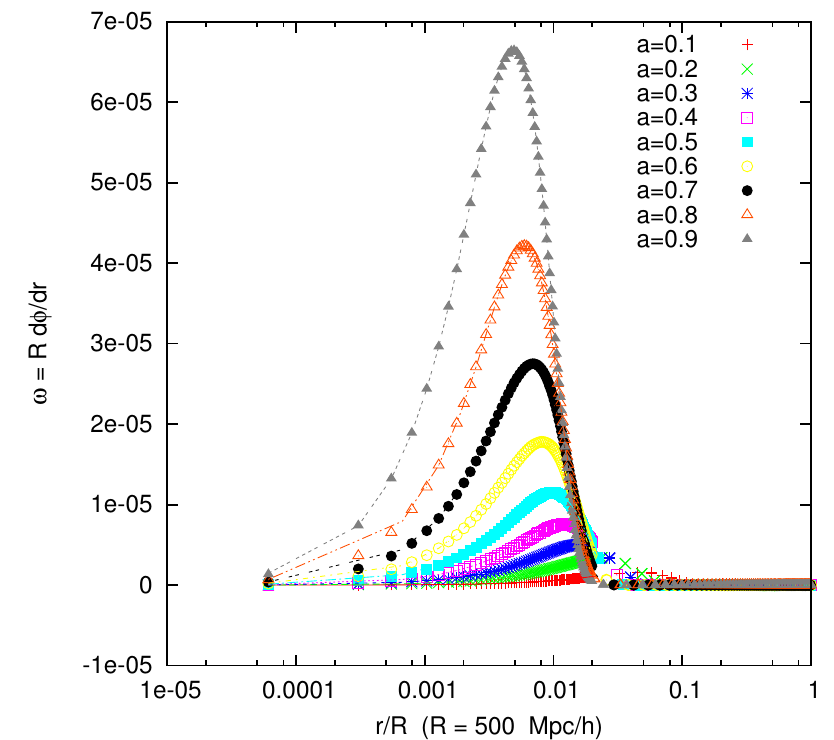}
\includegraphics[width=0.75\columnwidth]{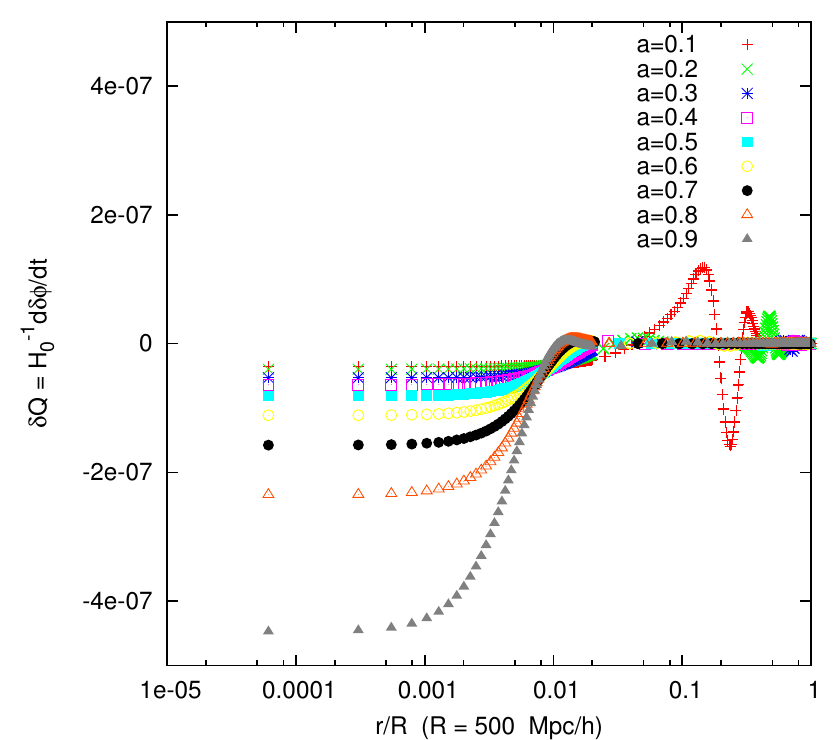}
\caption{The evolution of the Cubic Galileon field $\phi$ and its derivatives from $a=0.1$ to $a=1.0$ for a void with under-density $\delta_0 = -0.5$ today. The dashed lines in the $\omega$ plot shows the analytical quasi-static solution.}
\label{fig:void_evo}
\end{figure*}

\begin{figure*}
\includegraphics[width=0.75\columnwidth]{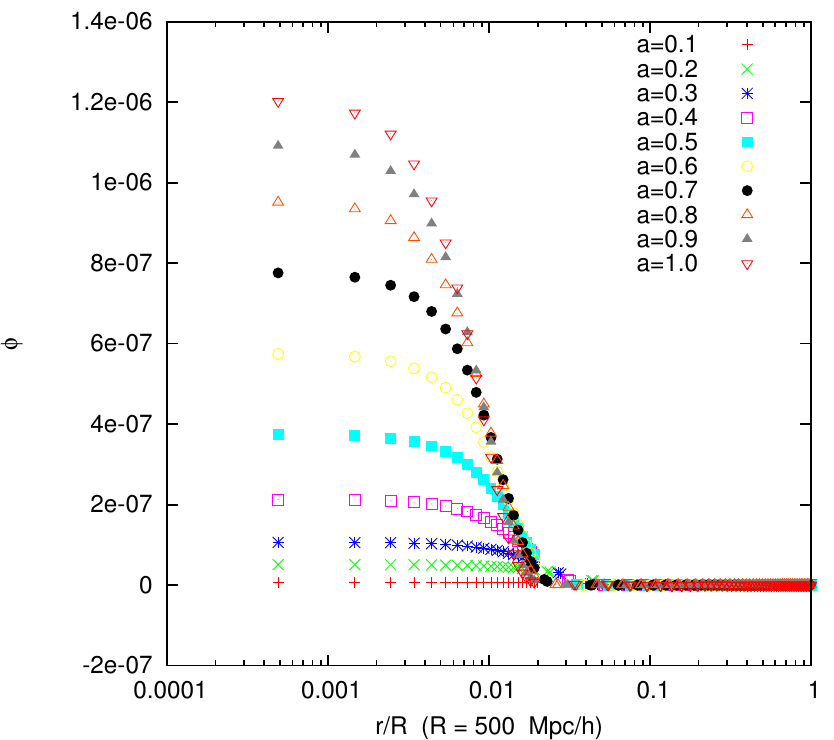}
\includegraphics[width=0.75\columnwidth]{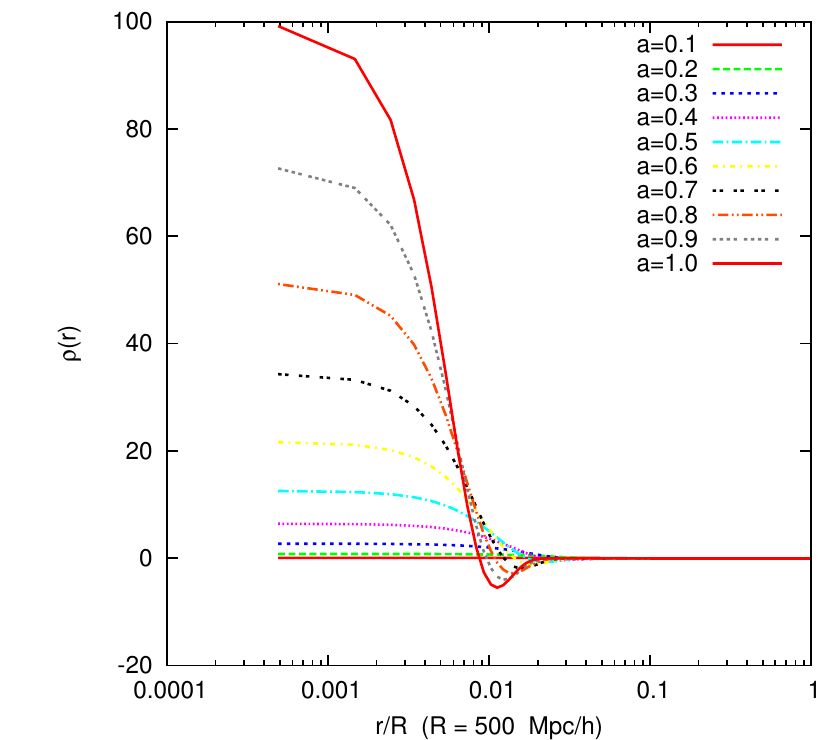}
\includegraphics[width=0.75\columnwidth]{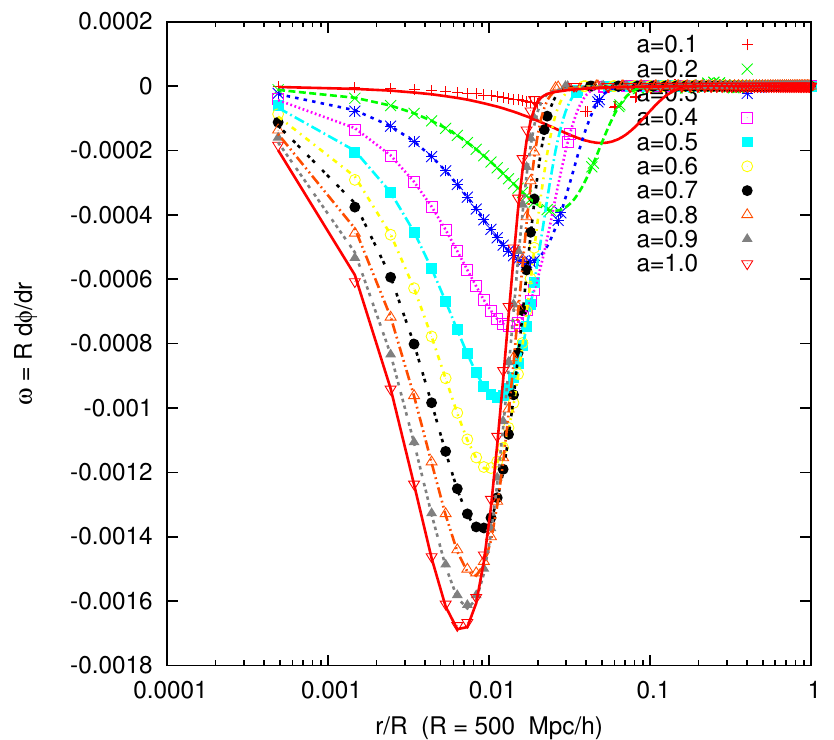}
\includegraphics[width=0.75\columnwidth]{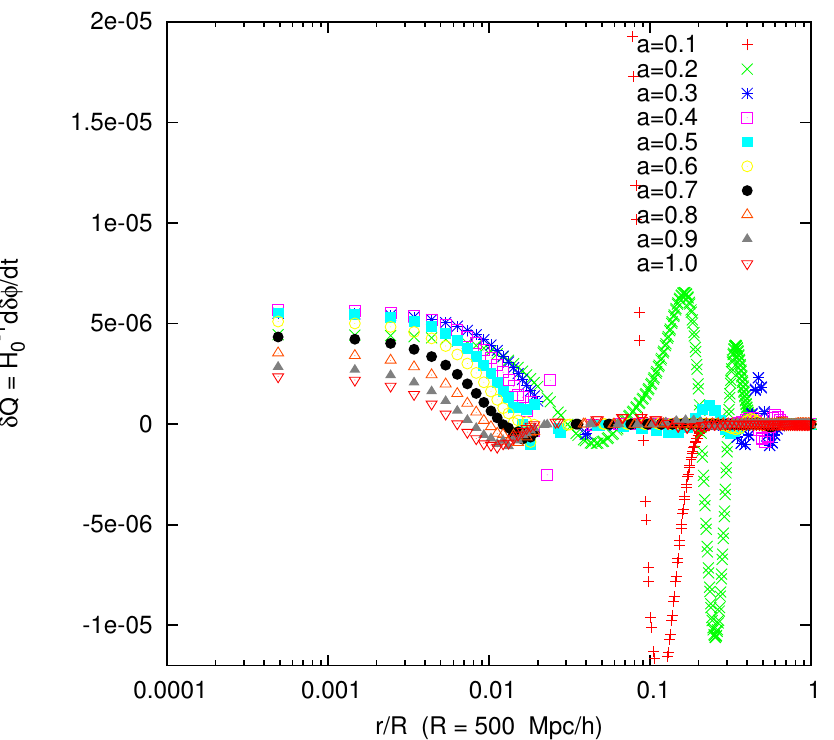}
\caption{The evolution of the Cubic Galileon scalar field $\phi$ and its derivatives from $a=0.1$ to $a=1.0$ for a cluster with over density $\delta_0 = 100$ today. The dashed lines in the $\omega$ plot shows the quasi-static solution.}
\label{fig:cluster_evo}
\end{figure*}

In all cases we looked at, for both over- and under-densities, the profile quickly (with the speed of sound) evolved towards the quasi-static solution and started oscillating around it until it finally settled down. In Fig.~(\ref{fig:void_evo}) we show the evolution as a function of time for the case of a growing void with density contrast $\delta_0 = - 0.5$ at $a=1.0$ and in Fig.~(\ref{fig:cluster_evo}) we show the evolution for a growing cluster with density contrast $\delta_0 = 100$ at $a=1.0$. 

We have also tried experiments with other initial conditions with the same result. The only case where we saw a real difference in the full solution compared to the quasi-static analytical approximation was when we considered the case of an object several giga-parsec large growing very rapidly in time. In that, unphysical, case the speed of sound of the field is not large enough for the field to be able to evolve quickly enough to catch up with the evolving quasi-static solution.

\begin{figure*}
\includegraphics[width=0.75\columnwidth]{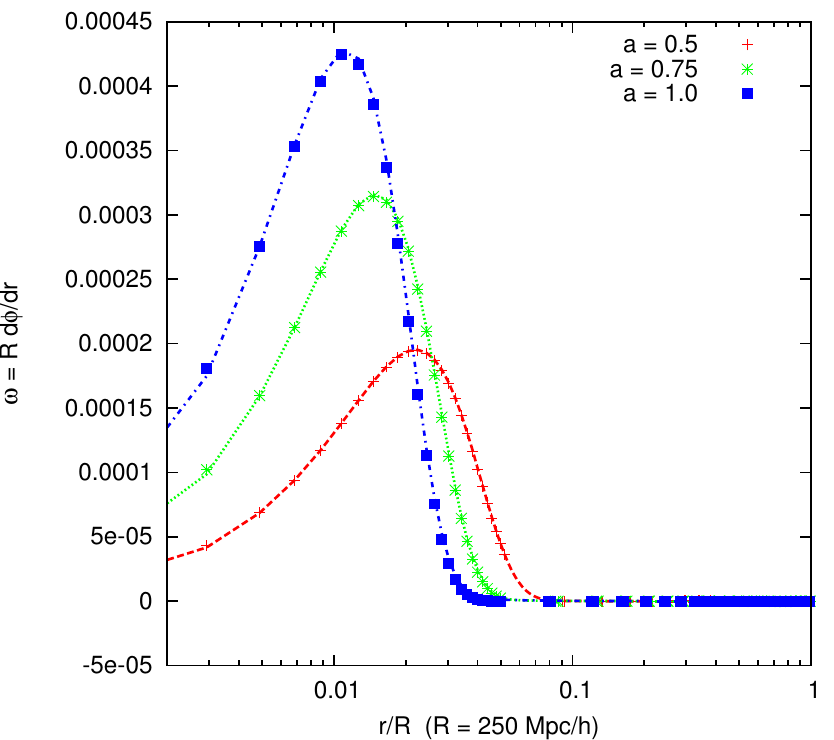}
\includegraphics[width=0.75\columnwidth]{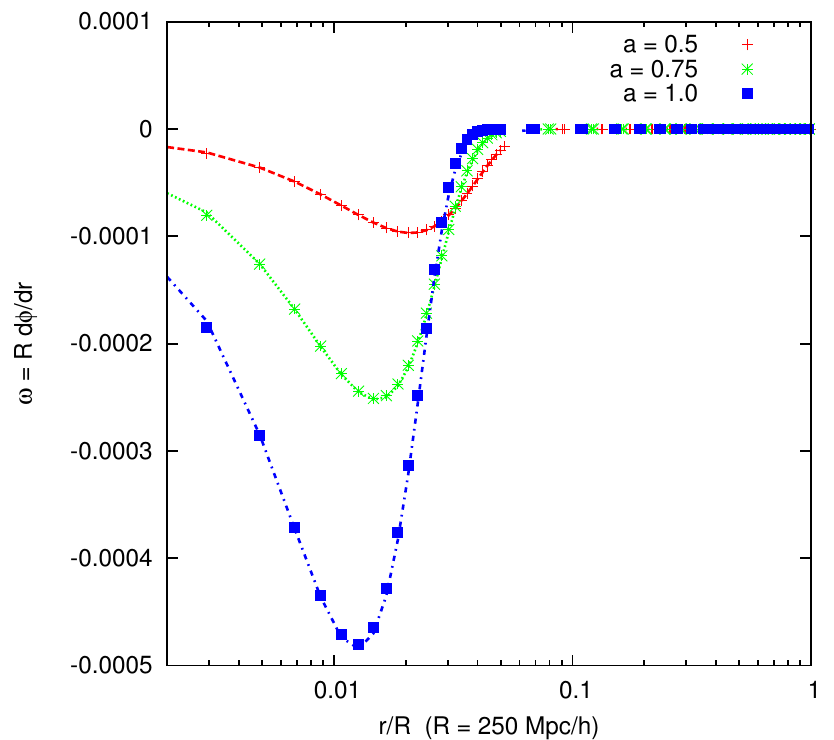}
\caption{The gradient of $\phi$, $\omega = R\frac{d\phi}{dr}$, at three different times for a cluster with $\delta_0 = 100$ at the present time ($\delta_m(r,a) \propto a^3$) for the normal branch (left) and the self-accelerating branch (right) of the DGP model. In both cases we have taken $r_CH_0 = 1.35$ which in the self-accelerating branch corresponds to $\Omega_m = 0.26$. The solid lines shows the analytic quasi-static solution.}
\label{fig:cluster_evo_dgp}
\end{figure*}

\begin{figure*}
\includegraphics[width=0.75\columnwidth]{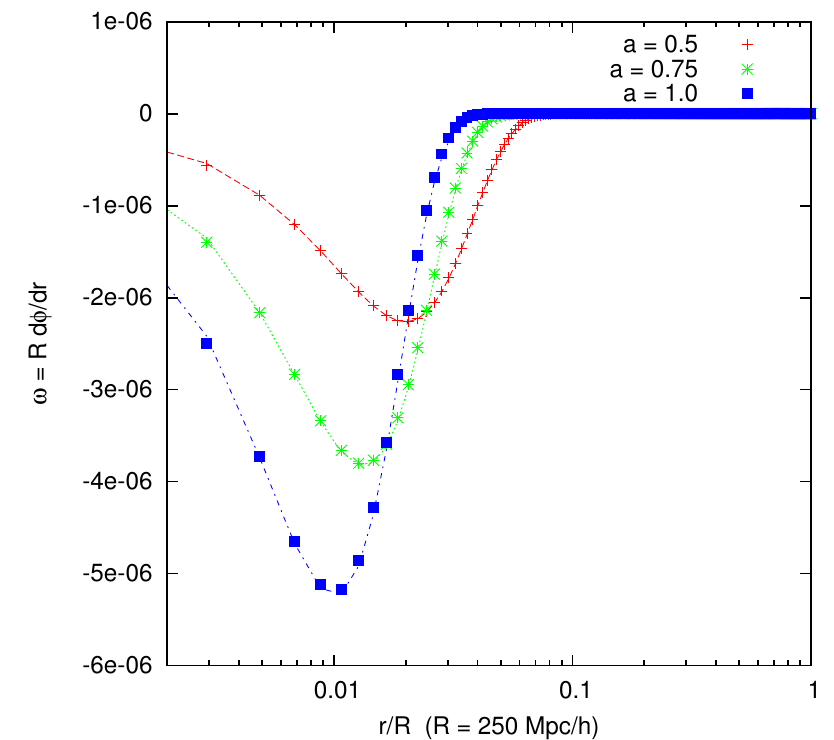}
\includegraphics[width=0.75\columnwidth]{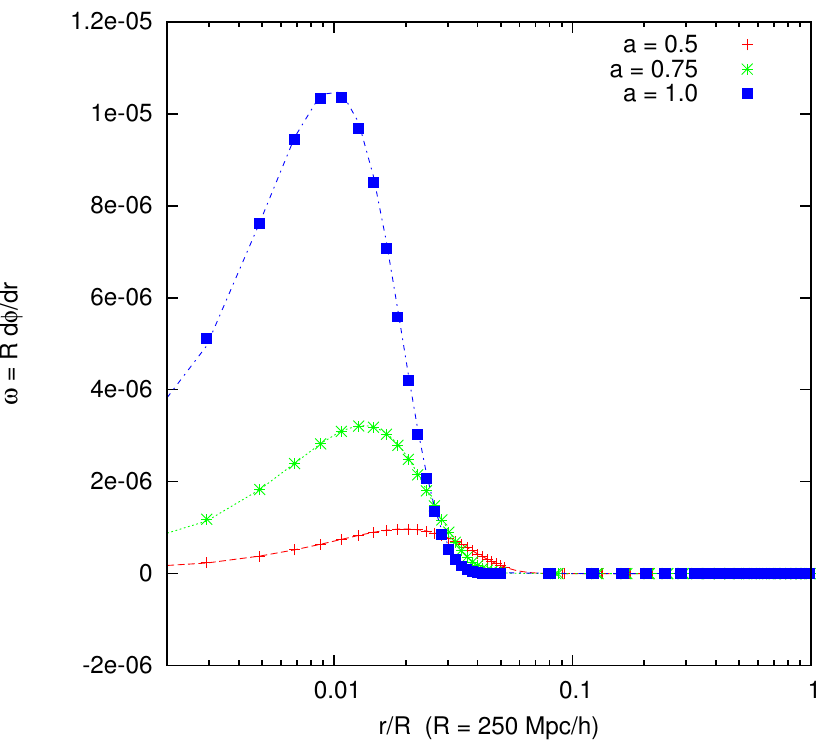}
\caption{The gradient of $\phi$, $\omega = R\frac{d\phi}{dr}$, at three different times for a void with $\delta_0 = -1.0$ at the present time ($\delta_m(r,a) \propto a^3$) for the normal branch (left) and the self-accelerating branch (right) of the DGP model. In both cases we have taken $r_CH_0 = 1.35$ which in the self-accelerating branch corresponds to $\Omega_m = 0.26$. The solid lines shows the analytic quasi-static solution.}
\label{fig:void_evo_dgp}
\end{figure*}

\subsection{Is the Vainshtein solution stable?}

To test the stability of the Vainshtein solution we run simulations where we have a fixed density profile where the scalar field has relaxed at the quasi-static solution and then we send in waves in the scalar field traveling towards the object. To form these waves we take advantage of reflecting boundary conditions to get the waves created when the scalar field evolved from $\phi=0$ and into the Vainshtein solution. By using a large enough box we can get these waves to reflect from the boundary and return to the center at any give time we want.

\begin{figure*}
\includegraphics[width=0.75\columnwidth]{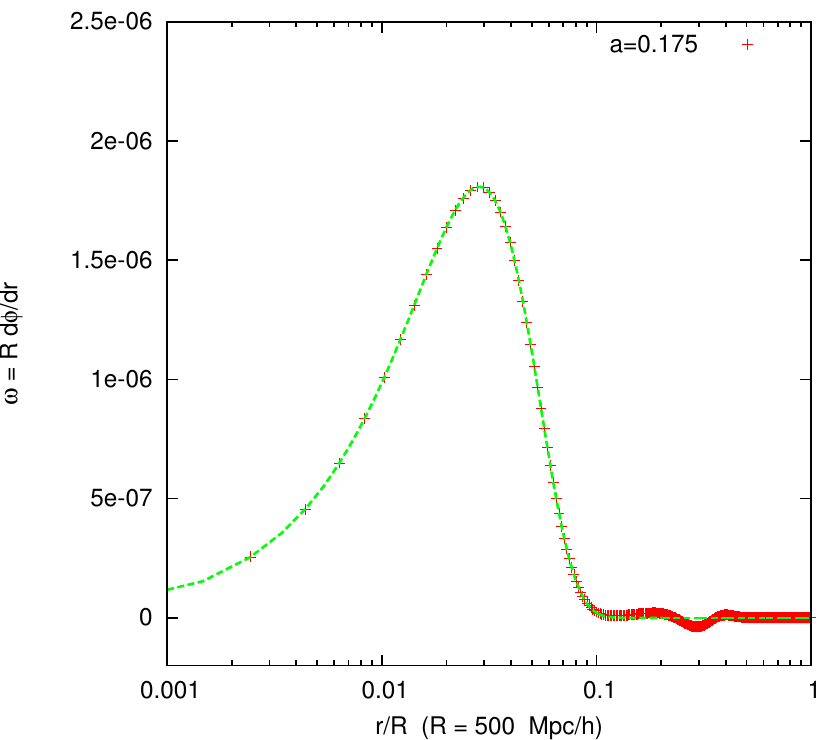}
\includegraphics[width=0.75\columnwidth]{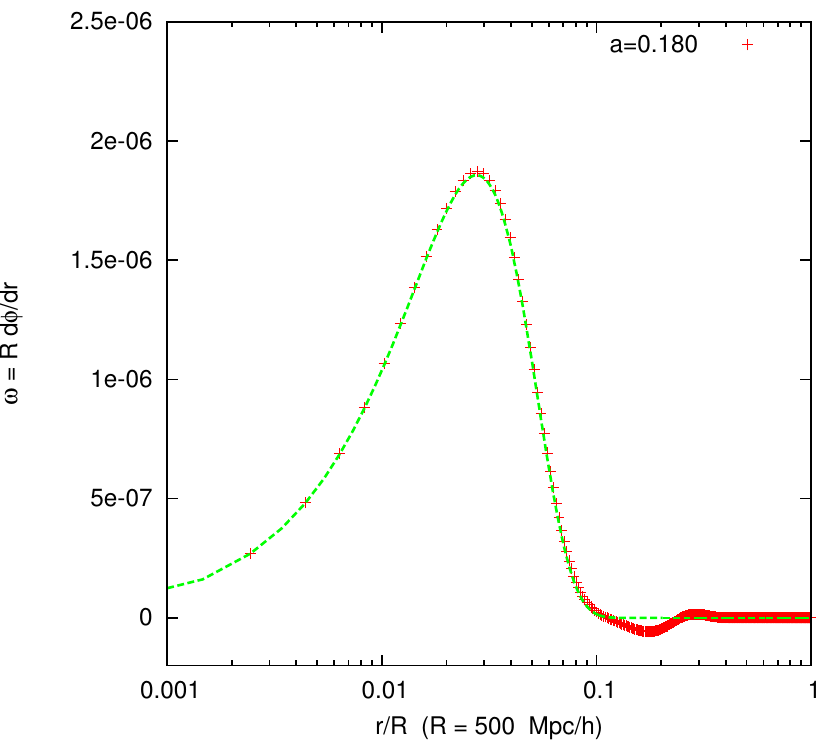}
\includegraphics[width=0.75\columnwidth]{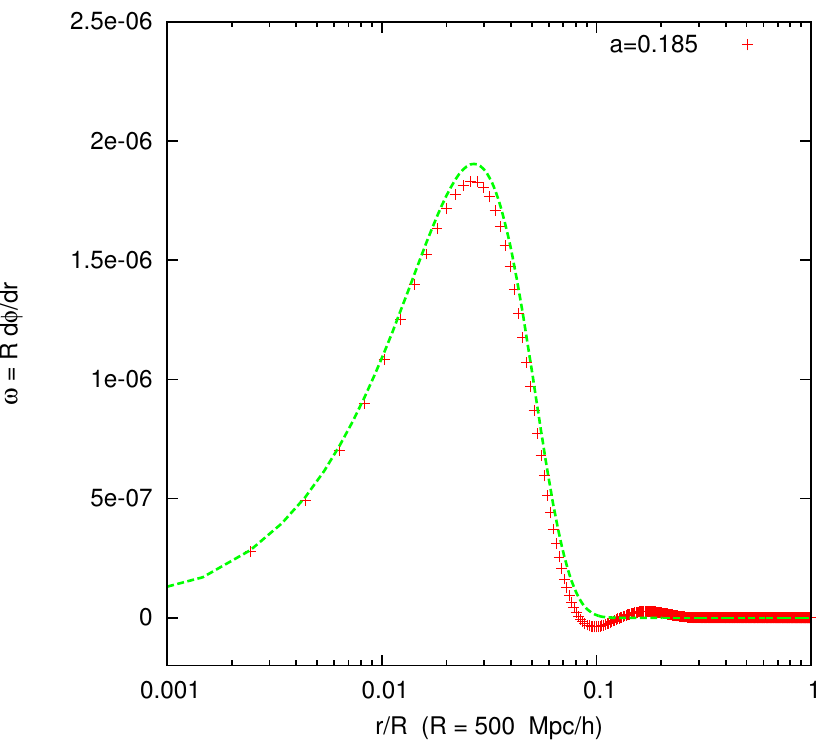}
\includegraphics[width=0.75\columnwidth]{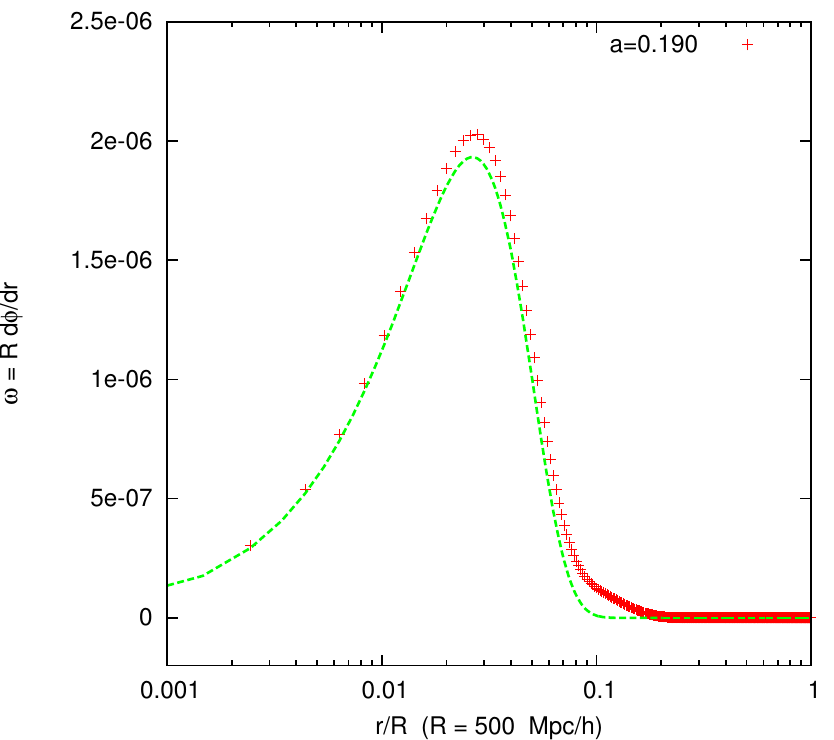}
\includegraphics[width=0.75\columnwidth]{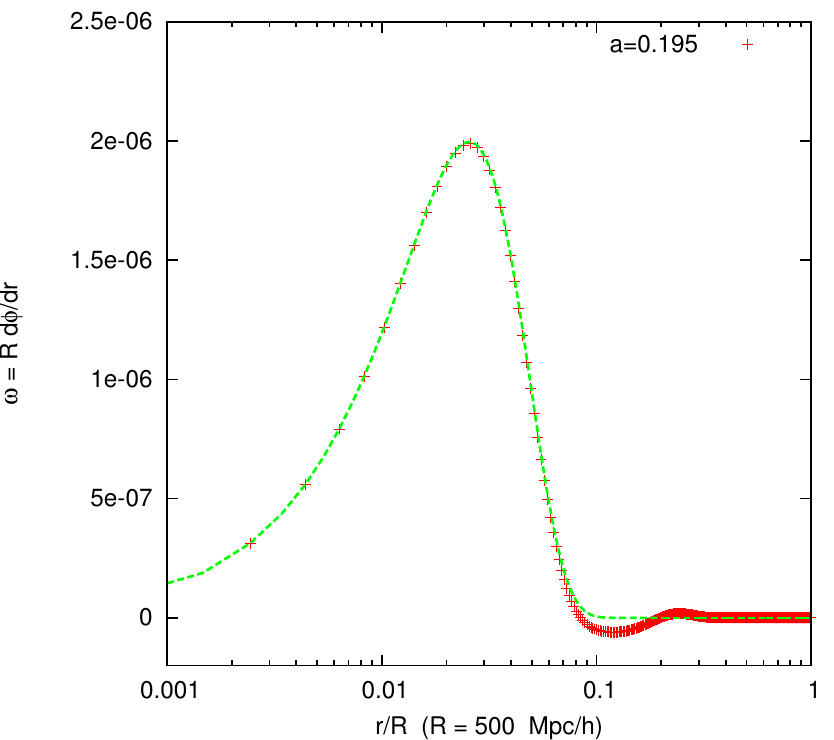}
\includegraphics[width=0.75\columnwidth]{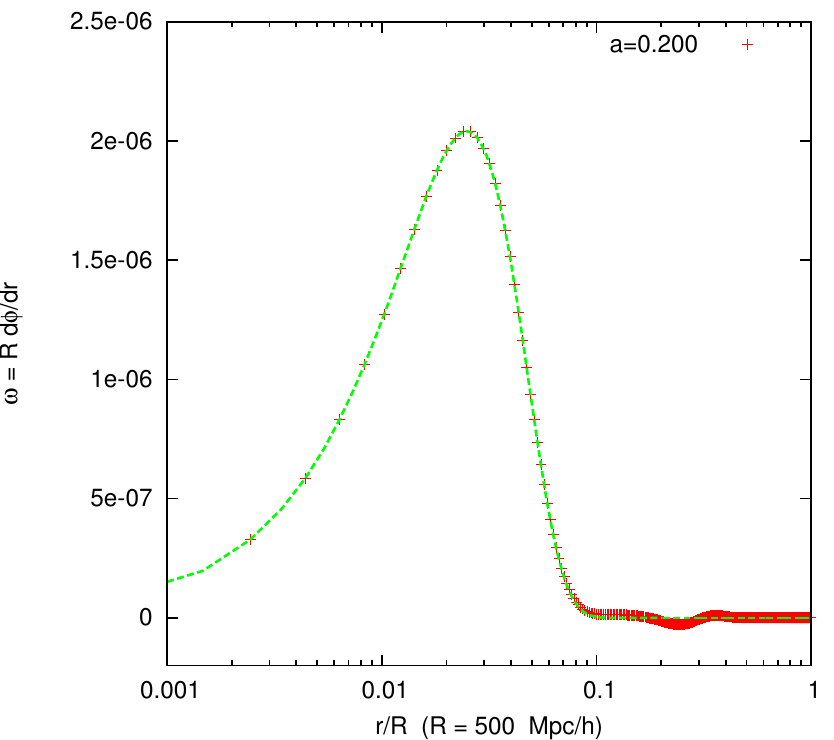}
\caption{We show the response of the $\omega$ profile to a scalar wave (coming in from the right). The wave reflects from the profile which then quickly returns to the quasi-static solution.}
\label{fig:stable}
\end{figure*}

When scalar waves hit the object they are quickly reflected back out again. The quasi-static solution remains stable under this process, changing very little, and only when we hit the profile with waves of very large amplitude are we able to significantly change the profile for a little while. However after the waves have been reflected the field profile relaxes to its old position, see Fig.~(\ref{fig:stable}). The Vainshtein solution seems very stable to interference from incoming scalar waves.

It would be interesting to redo this experiment for the case where the matter making up the density profile actually experiences the fifth-force present and study if scalar waves can have any impact on the density distribution. This is however beyond the reach of this paper.

\subsection{How good is the quasi-static approximation?}

To start with the punchline: the quasi-static approximation is excellent in almost all cases we have checked. If we start the simulation with $\phi$ being away from the quasi-static solution then it will quickly start to evolve towards it and once it has been reached, it undergoes damped oscillations until it quickly settles down to it (as shown in the previous section). If the quasi-static solution is evolving (like it will be if our density profile is growing/decaying with time) then we find that the quasi-static solution acts as an attractor which the full numerical solution follows closely.

In Fig.~(\ref{fig:void_evo}) and Fig.~(\ref{fig:cluster_evo}) (see the lower left panel) we show the evolution of $\omega$ together with the quasi-static solution for the ubic Galileon for a void and a cluster respectively. The same figure for the DGP model can be seen in Fig.~(\ref{fig:void_evo_dgp}) and Fig.~(\ref{fig:cluster_evo_dgp}).

The deviations from quasi-static evolution we find for the DGP model are consistent with the simulations of \cite{2009PhRvD..80d3001S} where the effect of the approximation was estimated from the time-evolving quasi-static solution (but without actually having any time evolution in the simulation itself).

It would be interesting to study what happens when the density profile changes dramatically in a short period of time, but this is beyond what we aim to study in this paper.

\subsection{Validity of other approximations}

We have run simulations where we evolve the field equation both with and without taking the metric derivative terms $\dot{\Phi},\dot{\Psi},d\Phi/dr$ and $d\Psi/dr$ into account in the field equation. Note that neglecting terms like $\dot{\Phi},\dot{\Psi}$ are strictly speaking part of the usual quasi-static approximation, but we have chosen to study it separately here.

\begin{figure*}
\includegraphics[width=0.75\columnwidth]{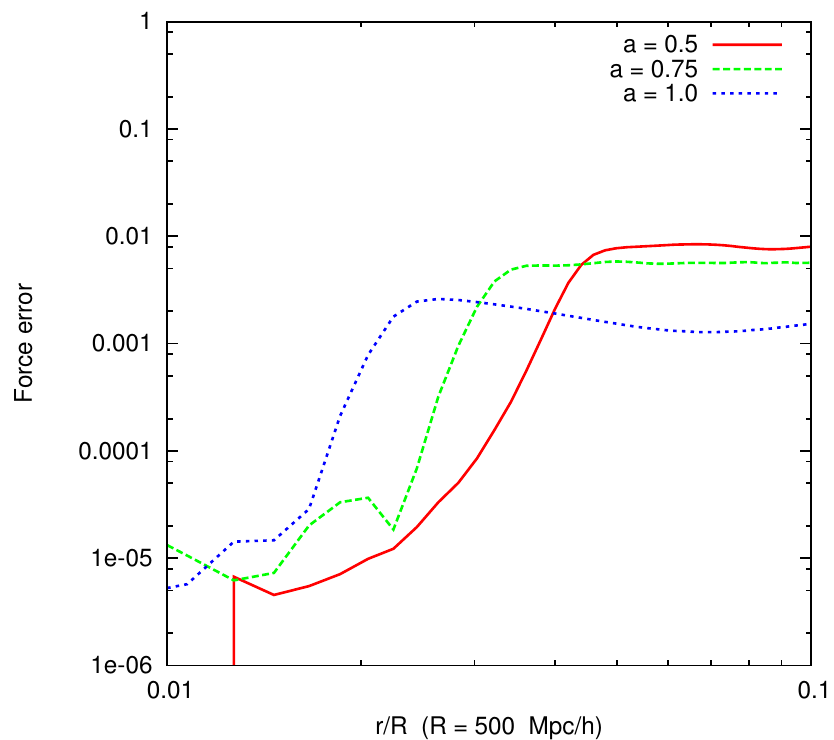}
\includegraphics[width=0.75\columnwidth]{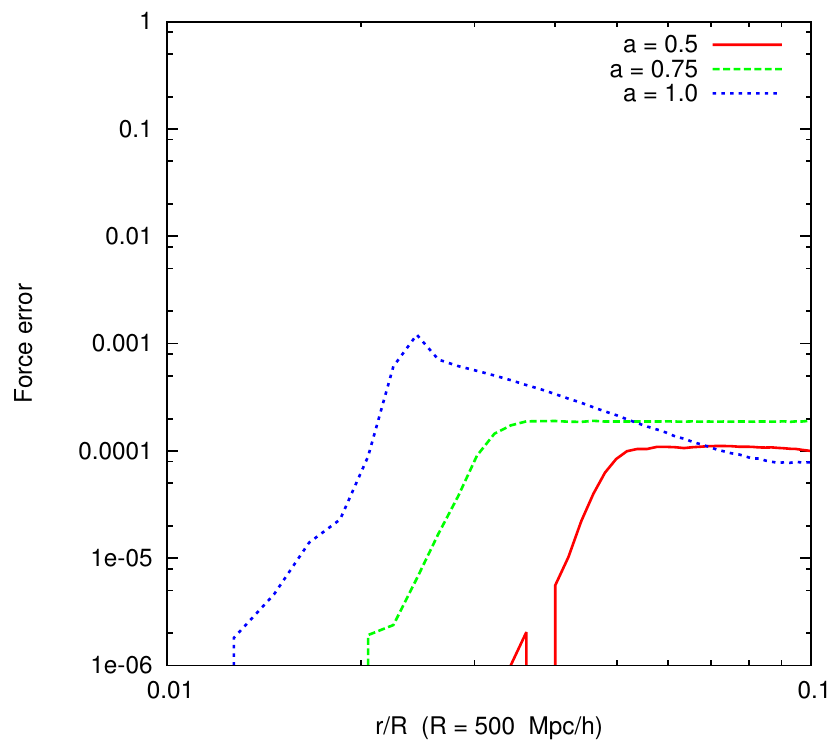}
\caption{The difference in the force $\omega = R\frac{d\phi}{dr}$ for a simulation where we included metric derivatives ($\dot{\Phi},d\Phi/dr$ etc.) compared to one where we neglected these. The simulations are of a cluster with overdensity $\delta_0 = 100$ (left) and a void of underdensity $\delta_0 = -0.5$ (right) at the present time.}
\label{fig:Phi_error}
\end{figure*}

In the equation of motion we do enforce this approximation by simply setting all the $\mathcal{A}$ and $\mathcal{B}$ terms in Eq.~(\ref{eq:eom_code_units} and Eq.~(\ref{btermsdgp}) to zero. The difference in the fifth-force $\omega = R\frac{d\phi}{dr}$ between the two simulations, with and without this approximation, is shown in Fig.~(\ref{fig:Phi_error}).

The difference in the force inside the overdensity is seen to be at the $10^{-5}$ level which is of the same order of magnitude as the Newtonian potential $\Psi$ of the object in agreement with the rough arguments we gave in Sec.~(\ref{sect:smallterms}). Outside the overdensity, on the other hand, the difference can be as large as $10^{-2}$. However we should note that in the region ($r \gtrsim 10$ Mpc/h from the center of the overdensity) where we find the largest difference, the force itself is very close to zero, see Fig.~(\ref{fig:cluster_evo}), so the physical effect this difference would have on matter located in this region is practically as small as inside the object. For the underdensity we simulated we get a very similar result.

\subsection{Breakdown of the numerical solution in deep voids}

Lastly we turn to the case discussed in Sec.~(\ref{sect:breakdown}). For the Cubic Galileon, at late times and in deep voids the quasi-static approximation ceases to exist, i.e. we get complex solutions in Eg.~(\ref{vsteinsol}). When we reach the regime where the quasi-static approximation breaks down, so do our simulations. In particular, we find that the denominator $B$ in the field equation $\frac{dQ}{dt} = \frac{A}{B}$ crosses zero, leading to nonsensical solutions.  If $B\to 0$ then $Q \to \infty$ and the speed of sound also diverges, see Eg.~(\ref{eq:eom_code_units}), unless $A\to 0$ at the same time. In practice we stopped the simulation if $B$ crossed zero, and if we continue the simulation past this point then the profile quickly develops kinks and breaks down (in this regime the sound speed squared is negative). This effect happens in our simulations no matter how small we take the time-steps to be. We have gone down to time-steps of the order $\delta a \sim 10^{-10}$ corresponding to 1 week (!) of cosmic time per time-step and still the solution crosses zero without any sign of turning around. This problem is encountered for both methods of integration and with all gridsizes, from $N=100$ to $N=10000$, we have tried. 

There seems to be no terms in the equation of motion that are able to stop $B$ from crossing zero and even though one should be careful in extracting definite conclusions from a failed numerical experiment we think it is, at least, very unlikely that the quasi-static approximation is able to alleviate this problem and that it is indeed a true instabillity of the Cubic Galileon model. 

From a more theoretical point of view, a similar type of instabillity was discussed in \cite{2011JCAP...11..021E} for the G-bounce scenario.

For the DGP model we do not have this problem and both the numerical solution and the quasi-static solution exists for all voids at all times.

\section{Summary}

We have studied the Vainshtein mechanism without restricting ourselves to the quasi-static approximation. By solving for the full time-evolution of the DGP and the Cubic Galileon scalar field in a spherical symmetric spacetime we are able to demonstrate that the quasi-static approximation is indeed a very good approximation. This result has previously been explicitly shown for several other modified gravity models and with this study there are now explicit checks of this approximation for all the major screening mechanisms present in the literature.

We found that the  quasi-static Vainshtein solution is a stable attractor for the evolution of the scalar field. It is an attractor in the sense that even if we release the scalar field far away from this solution it will quickly evolve into and relax to this solution. It is stable in the sense that it is hard to disrupt it: incoming scalar waves are absorbed by the profile and quickly emitted back again without significantly changing the profile.

We have also demonstrated, as expected, that other approximations closely linked to the quasi-static approximation are also valid. For example we found that neglecting terms proportional to time and single spatial derivatives of the metric potentials ($\dot{\Phi},d\Phi/dr$) in the field equation for $\phi$ have very little impact on the solution.

The final question we set out to answer was if relaxing the quasi-static approximation could aleviate the problem found in \cite{2013JCAP...10..027B} regarding the breakdown of the solution in deep voids close to the present time. By solving for the full evolution of the scalar field we encountered exactly the same problem. Our numerical solution breaks down at roughly the same point in time independent of the time-steps, gridsize and numerical integrator we choose. It therefore seems very likely that we are dealing with a true instabillity of the model.

\section*{Acknowledgements}

We thank Alexandre Barreira and Ignacy Sawicki for useful discussions. PGF and HAW are supported by STFC, BIPAC and the Oxford Martin School. The calculations for this paper were performed on the DiRAC Facility jointly funded by STFC and the Large Facilities Capital Fund of BIS.

\appendix

\section{The $\mathcal{A}$ terms}\label{app:aterm}
Here we show the expressions for the $\mathcal{A}_{1-6}$ factors in the equation of motion Eq.~(\ref{eq:eom_code_units}):
\begin{widetext}
\begin{align}
\mathcal{A}_1 &= \left[18EQ^2 - 2E\sigma^2\omega^2 + \frac{Qc_2}{c_3} - 4Q\sigma^2\nabla_y^2\phi\right]\\
\mathcal{A}_2 &= \left[\left(\frac{dE}{dx}\right)(6Q^2-2\sigma^2\omega^2) + 36EQ^2 - 4E\sigma^2\omega^2 + \frac{3Qc_2}{c_3} - 8\sigma^2\omega\left(\frac{dQ}{dy}\right) - 8Q\sigma^2\nabla_y^2\phi\right]\\
\mathcal{A}_3 &= \left[-4EQ\sigma\omega + \frac{4\sigma^3\omega^2}{y} + \frac{\sigma \omega c_2}{c_3} + 8Q\sigma \left(\frac{dQ}{dy}\right) - 4\sigma^3\omega\nabla_y^2\phi\right]\\
\mathcal{A}_4 &=  \left[-8EQ\sigma\omega - \frac{20\sigma^3\omega^2}{y} - \frac{\sigma \omega c_2}{c_3} + 8\sigma^3\omega\nabla_y^2\phi\right]\\
\mathcal{A}_5 &= \left[-8Q\sigma \omega\right]\\
\mathcal{A}_6 &=  \left[6Q^2 - 2\sigma^2\omega^2\right]
\end{align}
\end{widetext}

\section{Equations for the DGP model}\label{app:dgp}
Here we present, for completeness, the DGP equations
\begin{widetext}
\begin{align}\label{eq:dgpfull}
&H\frac{dQ}{dy}\left[1 + 4\tilde{r}\sigma^2\nabla_y^2\phi - 12\tilde{r}EQ -4EQ \tilde{r}\frac{d\Phi}{dx} - 8EQ\tilde{r}\frac{d\Psi}{dx} + 8\sigma^2\tilde{r}\omega\frac{d\Psi}{dy} - 8\sigma^2\tilde{r}\omega\frac{d\Phi}{dy} \right] = -3EQ + \sigma^2\nabla_y^2\phi\left[1 - 8EQ\tilde{r} + 8\sigma^2\tilde{r} \frac{\omega}{y}\right]\nonumber\\ 
&- \frac{\Omega_m \delta_m}{a^3\beta}+ \tilde{r}\left[12E^2Q^2+4\sigma^2\left(\frac{dQ}{dy}\right)^2 - 8E\sigma^2\omega\frac{dQ}{dy} + 4E^2\sigma^2 \omega^2 - \frac{12\sigma^4\omega^2}{y^2}\right] + E\mathcal{B}_1 \frac{d\Phi}{dx} + E\mathcal{B}_2 \frac{d\Psi}{dx} + \sigma\mathcal{B}_3 \frac{d\Phi}{dy} + \sigma\mathcal{B}_4 \frac{d\Psi}{dy}
\end{align}
where $\tilde{r} = \frac{(r_cH_0)^2}{6\beta(a)}$. The $\mathcal{B}$ terms are
\begin{align}\label{btermsdgp}
\mathcal{B}_1 &= Q - \tilde{r}\left[8EQ^2  - \frac{8Q\sigma^2\omega}{y} - 8\sigma^2 \omega\frac{dQ}{dy} + 8E\sigma^2\omega^2\right]\\
\mathcal{B}_2 &= 3Q - \tilde{r}\left[28EQ^2 -12Q\sigma^2\nabla_y^2\phi + \frac{8Q\sigma^2\omega}{y} \right]
 \\
\mathcal{B}_3 &= \sigma\omega - \tilde{r}\left[ 8EQ\sigma\omega - \frac{8\sigma^3\omega^2}{y} \right]\\
\mathcal{B}_4 &=  -\sigma\omega - \tilde{r}\left[ 8Q\sigma \frac{dQ}{dy} - 12EQ\sigma\omega +4\sigma^3\omega\nabla_y^2\phi\right]
\end{align}
\end{widetext}
In the DGP model the background value for the scalar field is simply $\overline{Q} = 0$. The Hubble equation for the self-accelerating branch is given by
\begin{align}
E(a) = \frac{1}{2r_cH_0} + \sqrt{\Omega_m a^{-3} + \frac{1}{(2r_cH_0)^2}}
\end{align}
where $\Omega_m = 1-\frac{1}{r_cH_0}$ or equivalently $r_cH_0 = \frac{1}{1-\Omega_m}$. For the normal branch we assumed dark-energy in such a form to give us a $\Lambda$CDM background
\begin{align}
E(a) = \sqrt{\Omega_m a^{-3} + \Omega_\Lambda}
\end{align}
where $\Omega_\Lambda = 1 - \Omega_m$.

The linearized equation of motion reads
\begin{align}
E(a)\frac{d\delta Q}{dx} - \sigma^2\nabla_y^2\delta\phi + 3E(a)\delta Q - \frac{\Omega_m \delta_m}{a^2\beta(a)} = 0
\end{align}
where $\delta Q = \frac{1}{H_0}\frac{d\delta\phi}{dt}$.

\section{Useful equations}\label{app:useful}
In this appendix we present some very useful equations needed to derive the field equation for a scalar field theory for spherical symmetry and in a perturbed, to first order in perturbation theory for the metric potentials, FRLW metric in the Newtonian gauge:
\begin{align}
ds^2 = -(1+2\Psi)dt^2 + a^2(1-2\Phi)\times\nonumber\\(dr^2 + r^2d\theta^2 + r^2\sin^2\theta d\phi^2)
\end{align}
Below we use the notation $\{1,2,3,4\} = \{t,r,\theta,\phi\}$, a dot is a time-derivative and a prime is a radial derivative $'\equiv d/dr$. To simplify the notation we use $\nabla^2$ for the radial Laplace operator: $\nabla^2\phi \equiv \phi'' + \frac{2}{r}\phi'$.

We have the following relations for the Christoffel symbols. The $\Gamma^{1}$ terms reads
\begin{align}
\Gamma_{11}^{1} &= \dot{\Psi}\\
\Gamma_{12}^{1} &= \Psi'\\
\Gamma_{22}^{1} &= Ha^2 - 2Ha^2(\Phi+\Psi) - a^2\dot{\Phi}\\
\Gamma_{33}^{1} &= r^2a^2H - 2r^2a^2 H(\Phi +\Psi) - r^2a^2\dot{\Phi}\\
\Gamma_{44}^{1} &= r^2a^2H\sin^2\theta - r^2a^2\sin^2\theta(4H + \dot{\Psi})
\end{align}
The $\Gamma^{2}$ terms reads
\begin{align}
\Gamma_{11}^{2} &= \frac{\Psi'}{a^2}\\
\Gamma_{12}^{2} &= H - \dot{\Phi}\\
\Gamma_{22}^{2} &= -\Phi'\\
\Gamma_{33}^{2} &= -r + r^2\Phi'\\
\Gamma_{44}^{2} &= -r\sin^2\theta - 2r\sin^2\theta(\Phi-\Psi) + r^2\sin^2\theta\Psi'
\end{align}
The $\Gamma^{3}$ terms reads
\begin{align}
\Gamma_{13}^{3} &= H - \dot{\Phi}\\
\Gamma_{23}^{3} &= \frac{1}{r} - \Phi'\\
\Gamma_{44}^{3} &= -\frac{\sin(2\theta)}{2} - (\Phi-\Psi)\sin(2\theta)
\end{align}
The $\Gamma^{4}$ terms reads
\begin{align}
\Gamma_{14}^{4} &= H - \dot{\Psi}\\
\Gamma_{24}^{4} &= \frac{1}{r} - \Psi'\\
\Gamma_{34}^{4} &= \cot\theta
\end{align}
The rest of the non-zero terms follows from using the symmetry $\Gamma^{\alpha}_{\mu\nu}=\Gamma^{\alpha}_{\nu\mu}$ on the terms presented above.

The determinant of the metric $g$ is
\begin{align}
\det{g} = -r^4 a^6\sin^2\theta(1-4\Phi)
\end{align}
Some useful components of the Ricci tensor are
\begin{align}
R_{11} &= 3\frac{\ddot{a}}{a} - \frac{1}{a^2}\nabla^2\Psi - 4H\dot{\Phi} - 5H\dot{\Psi} - 2\ddot{\Phi} - \ddot{\Psi}\\
R_{22} &= -2a^2H^2 - a^2 \frac{\ddot{a}}{a}(1-2\Phi-2\Psi) + 4a^2H^2(\Phi+\Psi)\nonumber\\ &- \nabla^2\Phi + 5a^2H\dot{\Phi} + 2a^2H\dot{\Psi} + a^2\ddot{\Phi}\\
R_{12} &= -2H\Psi' + \frac{\dot{\Phi}-\dot{\Psi}}{r} - \dot{\Phi}'-\dot{\Psi}'
\end{align}
The Ricci scalar becomes
\begin{align}
R = &-6\left[H^2 + \frac{\ddot{a}}{a}\right](1-2\Psi) - \frac{2}{a^2}\nabla^2\Phi \nonumber\\
&+ \frac{2}{ra^2}(\Phi' - \Psi') + 16H\dot{\Phi} + 14H\dot{\Psi} + 4\ddot{\Phi} + 2\ddot{\Psi}
\end{align}
Some useful derivatives of the scalar field is
\begin{align}
g^{\mu\nu}\phi_{,\mu}\phi_{,\nu} &= \frac{\phi'^2}{a^2}(1+2\Phi) - \dot{\phi}^2(1-2\Psi)\\
\square\phi &= \frac{(1+2\Phi)}{a^2}\nabla^2\phi - 3H\dot{\phi}(1-2\Psi) - \ddot{\phi}
\end{align}

\newpage
{}
\newpage
{}
\newpage
{}

\end{document}